%% file: main.tex
\documentclass[letter,11pt]{article}
\usepackage[top=1in,bottom=1in,left=1in,right=1in]{geometry}
\usepackage{slashed}
\usepackage{epsfig}
\usepackage{comment}
\usepackage{cancel}
\usepackage{bbm}
\usepackage{array}
\usepackage{bigints}
\usepackage{booktabs}
\usepackage{color}
\usepackage{dsfont}
\usepackage{float}
\usepackage{framed}
\usepackage{graphicx}
\usepackage{indentfirst}
\usepackage{mathrsfs}
\usepackage{multirow}
\usepackage{subdepth}
\usepackage{titlesec}
\usepackage[dotinlabels]{titletoc}
\usepackage{wrapfig}
\usepackage[all]{xy}
\usepackage[vcentermath]{youngtab}
\usepackage{relsize}
\usepackage{hyperref}
\numberwithin{equation}{section}

\usepackage{tikz}
\usepackage{soul}

\newcommand\EatDot[1]{}

\hypersetup{pdfauthor={Name}} 
\usepackage[utf8]{inputenc}
\usepackage{slashed}
\usepackage{amsmath}
\usepackage{amsfonts}
\usepackage{amssymb}
\usepackage{cite}
\usepackage{mathtools}

\usepackage{cleveref}
\crefname{equation}{Eq.}{Eqs.} 
\crefrangelabelformat{equation}{(#3#1#4--#5#2#6)}
\crefname{section}{§}{§§}
\Crefname{section}{§}{§§}

\usetikzlibrary{shadings}
\usepackage{setspace}

\def\half{{1\over 2}}

 \def\p{\partial}

\def\0{{(0)}}
\def\1{{(1)}}
\def\2{{(2)}}

\def\<{\langle }
\def\>{\rangle }

\newcommand{\bea}{\begin{eqnarray}}
\newcommand{\eea}{\end{eqnarray}}
\newcommand{\be}{\begin{equation}}
\newcommand{\ee}{\end{equation}}
\newcommand{\ba}{\begin{align}}
\newcommand{\ea}{\end{align}}

   \makeatletter
  \let\over=\@@over \let\overwithdelims=\@@overwithdelims
  \let\atop=\@@atop \let\atopwithdelims=\@@atopwithdelims
  \let\above=\@@above \let\abovewithdelims=\@@abovewithdelims
\renewcommand\section{\@startsection {section}{1}{\z@}%
                                   {-3.5ex \@plus -1ex \@minus -.2ex}
                                   {2.3ex \@plus.2ex}%
                                   {\normalfont\large\bfseries}}

\renewcommand\subsection{\@startsection{subsection}{2}{\z@}%
                                     {-3.25ex\@plus -1ex \@minus -.2ex}%
                                     {1.5ex \@plus .2ex}%
                                     {\normalfont\bfseries}}

\newcommand{\beq}{\begin{equation}}
\newcommand{\eeq}{\end{equation}}
\newcommand{\beqa}{\begin{eqnarray}}
\newcommand{\eeqa}{\end{eqnarray}}
\newcommand{\beqar}{\begin{eqnarray*}}

\def\[{\big[}
\def\]{\big]}

\def\ba{{\bar a}}

\def\be{{\bar \epsilon}}

\def\half{{1\over 2}}

\def\+{{(+)}}
\def\-{{(-)}}
\def\0{{(0)}}
\def\1{{(1)}}
\def\2{{(2)}}
\def\3{{(3)}}

\def\be{\begin{equation}}
\def\ee{\end{equation}}

\newcommand{\phih}{\hat{\phi}}
\newcommand{\psih}{\hat{\psi}}
\newcommand{\xph}{{\hat{x}^+}}
\newcommand{\xmh}{{\hat{x}^-}}

\begin{document}
\begin{titlepage}
\unitlength = 1mm
\ \\
\vskip 3cm
\begin{center}
 
{\huge{{\textsc{Self-Dual Black Holes \\ in  Celestial Holography}}}}

\vspace{1.25cm}
Erin Crawley,$^{*}$ Alfredo Guevara,$^{*\mathsection\dagger}$ Elizabeth Himwich,$^{*\mathsection}$ and Andrew Strominger$^{*\mathsection}$

\vspace{.5cm}

$^*${\it  Center for the Fundamental Laws of Nature, Harvard University,
Cambridge, MA 02138}\\ 
$^\mathsection${\it Black Hole Initiative, Harvard University, Cambridge, MA 02138}
\\ 
$^\dagger${\it  Society of Fellows, Harvard University,
Cambridge, MA 02138}\\ 

\vspace{0.8cm}

\begin{abstract} 
	We construct two-dimensional quantum states associated to four-dimensional linearized rotating  self-dual black holes in $(2,2)$ signature Klein space. The states are comprised of  global conformal primaries circulating on the celestial torus, the Kleinian analog of the celestial sphere. By introducing a generalized tower of Goldstone operators we identify the states as coherent exponentiations carrying an infinite tower of ${\rm w}_{1+\infty}$  charges or soft hair. We relate our results to recent approaches to black hole scattering, including a connection to Wilson lines, $\mathcal{S}$-matrix results, and celestial holography in curved backgrounds.
\end{abstract}

\vspace{1.0cm}

\end{center}

\end{titlepage}

\pagestyle{empty}
\pagestyle{plain}

\def\vx{{\vec x}}
\def\p{\partial}
\def\po{$\cal P_O$}
\def\i{{\rm initial}}
\def\f{{\rm final}}

\pagenumbering{arabic}
 

\tableofcontents

\section{Introduction }

Over the last few years there has been uncanny success within the perturbative amplitudes program in describing the scattering or mergers of black holes \cite{Cheung:2018wkq,Bern:2019crd,Bern:2021dqo,Bjerrum-Bohr:2018xdl,Damour:2017zjx,DiVecchia:2020ymx,Aoude:2020onz,Mogull:2020sak,Kalin:2019rwq,Damour_2016,Cheung:2020gyp,Bern:2019nnu,Cristofoli:2019neg,  Bjerrum-Bohr:2019kec, Bjerrum-Bohr:2021din,Bjerrum-Bohr:2021vuf, DiVecchia:2021ndb,DiVecchia:2021bdo,Kosower:2018adc, Kalin:2019inp,Levi:2020kvb,Levi:2020uwu,Kalin:2020mvi, Kalin:2020fhe, Damour:2020tta,Damour:2019lcq, Maybee:2019jus,  Arkani-Hamed:2019ymq,Bern:2020buy,Chung:2019duq,Chung:2020rrz, Cachazo:2017jef,Guevara:2017csg,Guevara:2018wpp,Guevara:2019fsj,Goldberger:2020fot, Goldberger:2017vcg, Goldberger:2017ogt,Bautista:2019tdr, Bautista:2019evw,Herrmann:2021lqe, Herrmann:2021tct,Jakobsen:2021smu, Bern:2020uwk,Jakobsen:2021lvp,Mougiakakos:2021ckm,Cristofoli:2020uzm,Vines:2018gqi,Blumlein:2020znm,Adamo:2022ooq,Adamo:2022rob}. In some circumstances  exact results are obtainable, and there is even a construction of a pure\footnote{It is interesting that the required quantum state  is not mixed.} 4D quantum state that reproduces some scattering amplitudes off of a single black hole \cite{Monteiro:2020plf,Monteiro:2021ztt,Britto:2021pud,Cristofoli:2021jas,Adamo:2022ooq}. The exactness of some results in part stems from the  existence of the Kerr-Schild form of the metric in which the linearized metric perturbation is exact \cite{Luna:2015paa,Chong:2004hw}. Much of the analysis effectively employs analytic continuation from $(2,2)$ Klein space $\mathbb{K}^{2,2}$ (rather than $(0,4)$ Euclidean space) where massless particles can propagate and efficient on-shell methods are available. The geometry and causal structure of Kleinian black holes was studied in      \cite{Crawley:2021auj}, where among other things it was shown that the self-dual Kerr-Taub-NUT family of black holes parameterized by the spin $\vec{a}$ can all be obtained from the $\vec{a}=0$ (Taub-NUT) case by a simple translation ($\vec{x}\to\vec{x}-\vec{a}$ in a particular choice of coordinates).

Clearly these results indicate still further curious and surprising properties of black holes whose ultimate origin should be understood.  

The goal of this paper is to incorporate these results into the framework of celestial holography (see the recent reviews \cite{McLoughlin:2022ljp,Raclariu:2021zjz,Pasterski:2021rjz,Pasterski:2021raf} and references therein).  The motivations are both to better understand  properties of black holes from a new perspective and to develop/test  the celestial bulk-to-boundary dictionary beyond the known relations between scattering amplitudes. Our main result will be the explicit construction of a 2D state describing a 4D self-dual Kerr-Taub NUT black hole that both carries the requisite infinite tower of nonzero ${\rm w}_{1+\infty}$ charges and gives the correct expectation values for the 2D operators dual to bulk metric modes.  

We restrict here to a linearized analysis only. Fully non-linear results have been obtained in the amplitudes context, but going consistently beyond leading order for us would require significantly more work including computation of all the operator product expansions. However, the non-linear successes in related contexts suggest a fully non-linear analysis may also be possible here. 

We present  self-dual black holes  in terms of primary states circulating in  the Lorentzian celestial torus \cite{atanasov2021_22Scat} at the boundary of Klein space. Such circulating 2D states were referred to as \textit{$L$-primaries} in \cite{atanasov2021_22Scat}, and are  boost eigenstates transforming covariantly under the $SL(2,\mathbb{R})_L\times SL(2,\mathbb{R})_R$ Lorentz group of  $\mathbb{K}^{2,2}$. Our construction is based on decomposing the rotating black hole metric in terms of these $L$-primaries. The main idea has a close analog in Lorentzian signature, for which solutions of the wave equation (e.g. perturbations of Minkowski space) can be expanded in $SO(3)$ rotation eigenstates. In particular, the aforementioned coordinate shift $\vec{x}\to \vec{x}-\vec{a}$ generates a full tower of spherical harmonic $\ell$-eigenstates in Lorentzian signature: 
\begin{equation}\label{eq:Lorex}
    \frac{1}{|\vec{x}-\vec{a}|} = \sum_\ell \frac{a^\ell}{r^{\ell+1}} P_{\ell}(\cos\theta), 
\end{equation}
where $|\vec{a}| = a$, $|\vec{x}| = r$, $\theta$ is the angle between $\vec{x}$ and $\vec{a}$, and $P_{\ell}$ is the $\ell$-th Legendre polynomial. In the context of classical Kerr-Schild metrics, the above tower is loosely related to the intrinsic multipole moments of a Kerr black hole with spin $\vec{a}$ \cite{Harte:2016vwo}. Remarkably, in Kleinian signature, we demonstrate that the above can be made precise: the self-dual Kerr-Taub-NUT metric is the generating function of torus $L$-primaries, which are boost eigenstates of (half-)integer weight rather than the rotation eigenstates of Lorentzian signature.

A feature of Klein space is that boost eigenstates are created by insertions of scattering states on the celestial torus \cite{atanasov2021_22Scat}. Using this fact we show that the gravitational potential can be expanded in radiative ($\mathcal{S}$-matrix) modes, thus recovering the results from \cite{Crawley:2021auj,guevara2021reconstructing,Monteiro:2020plf,Monteiro:2021ztt} that connect scattering amplitudes to classical spacetimes. Then, to obtain the associated quantum state on the celestial torus, we construct the two-dimensional vacuum, together with a generalized tower of Goldstone modes associated one-to-one with $L$-primary insertions. Assuming quantization of the Taub-NUT charges \cite{bunster2006monopoles, argurio2009supersymmetry, Argurio_2009} leads us to construct a black hole state $| M,a\rangle$ as a coherent state of Goldstone modes. The  coefficients in  this state are  fixed in terms of  the 2D expectation value of different modes of the two-dimensional metric operator $\langle M,a|H^{k}_{m,n}|M,a\rangle  $.
We find that these modes correspond precisely to an infinite set of ${\rm w}_{1+\infty}$ charges carried by the black hole state, corresponding to  a tower of soft hair \cite{Hawking:2016sgy,Hawking:2016msc}.  

For self-dual black holes, the metric operator can be written as the sum over a tower of positive-helicity soft graviton operators.  We use commutation relations derived from the pairing of positive-helicity soft gravitons and their negative-helicity Goldstone partners to derive the ${\rm w}_{1+\infty}$ charges of the self-dual black hole.\footnote{Although we restrict to the self-dual case here,   it is straightforward to perform an analogous construction and derive the $\overline{\rm w}_{1+\infty}$ charges for the anti-self-dual case, instead using the commutation relations of negative-helicity soft gravitons and positive-helicity Goldstone modes. However, taking the perspective that such commutation relations could be derived from celestial operator product expansions (OPEs), a puzzle arises about how to consider the self-dual and anti-self-dual cases simultaneously. These cases correspond to OPEs derived respectively from holomorphic and anti-holomorphic collinear limits, which have not been understood simultaneously in the celestial theory.  We leave this interesting open problem for future study. }

This paper is organized as follows.  In Section 2, we demonstrate that the linearized classical Kleinian self-dual Kerr-Taub-NUT metrics have a simple expression as a sum of conformal primary wavefunctions on the celestial torus.  In Section 3, we then recover the classical self-dual Kerr-Taub-NUT metrics as the expectation value of a soft graviton operator in a particular state on the celestial torus. In Section 4, we connect the symmetries of the 4D metric with the charges of the 2D state under ${\rm w}_{1+\infty}$. In Section 5,  we connect the classical Section 2 with the usual amplitudes formalism and note the connection between our construction and bulk Wilson lines. We conclude with a discussion in Section 6. Appendices A and B include more details about classical Kleinian metrics and operators on the celestial torus, respectively, and Appendix C presents a connection of our work to the recent four-point amplitudes calculations in \cite{Gonzo:2022tjm}. 

While this work was in preparation, a similar construction of integer-mode pairings and Goldstone mode coherent states appeared in \cite{Freidel:2022skz}. We hope to understand the connection more explicitly in future work.

\section{Integral Conformal Primary Expansion of the Classical Metric}

In this section, we expand  linearized metrics for spinning self-dual black holes  in an integer basis of   conformal primary graviton wavefunctions.

We expand  $(2,2)$ signature (Kleinian) metrics in the gravitational constant $G_N$:
\begin{equation}\label{eq:dx22}
    ds^2 = g_{\mu \nu} dX^\mu dX^\nu ,  \qquad g_{\mu \nu} = \eta_{\mu\nu} + \kappa \,  h_{\mu \nu} + \mathcal{O}\left(G_N^2\right) ,
\end{equation}
where $G_N = \frac{\kappa^2}{32\pi}$ is Newton's constant and $h_{\mu\nu} \sim \mathcal{O}(\kappa)$. In  special cases of  Kerr-Schild or double Kerr-Schild solutions, the linear term in the expansion gives the exact metric \cite{xanthopoulos1978exact}. In Kleinian signature, metrics can be real and self-dual or anti-self-dual, respectively satisfying
\begin{equation}\label{eq:selfdualcondition}
R_{\mu\nu\rho\sigma} = \pm\frac{1}{2}\varepsilon_{\mu\nu\alpha\beta}R^{\alpha\beta}_{\ \ \ \rho\sigma}.
\end{equation} 
A general linearized metric can be decomposed into self-dual ($h^+_{\mu \nu}$) and anti-self-dual ($h^-_{\mu \nu}$) pieces 
\begin{equation}\label{eq:hmunu_SDplusASD}
    h_{\mu\nu} = h^+_{\mu \nu} + h^-_{\mu \nu}.  
\end{equation}
In this paper we will specialize to self-dual (Kerr-)Taub-NUT metrics, for which the  construction of quantum states in Section \ref{sec:quantum} simplifies because only $ h^+_{\mu \nu}$ appears. Kleinian self-dual solutions were studied in \cite{Crawley:2021auj} that correspond to a particular case of double Kerr-Schild metrics for which the Einstein equations linearize \cite{Luna:2015paa}. While we do not use Kerr-Schild gauge in this work, one can explicitly match our linearized results to those of the non-linear solution using a coordinate shift reminiscent of the Talbot shift \cite{Crawley:2021auj,Talbot1969} (see \eqref{eq:nonlineartrans} in Appendix \ref{app:KTN}).  

\subsection{Flat Klein Space and the Celestial Torus}
In this section, we review some useful results from \cite{atanasov2021_22Scat} pertaining to Klein space. We begin with flat Klein space in double polar $(r, \phi, q, \psi)$ coordinates 
\begin{equation}\label{eq:flatKlein}
    ds^2 = dr^2 + r^2 d\phi^2 - dq^2 -q^2d\psi^2,
\end{equation}
where $r,q > 0 $, $\phi \sim \phi +2\pi$, and $\psi \sim \psi +2\pi$. We take the following choice of rectangular coordinates:
\begin{equation}\label{eq:TXYZinrqphps}
    X^{\mu}=(T,X,Y,Z)= (q \sin \psi,r \cos\phi,r \sin\phi,q \cos \psi)\,,
\end{equation}
in which the metric becomes
\begin{equation} \label{eq:cartcoord}
    ds^2 = -dT^2 + dX^2 +dY^2 -dZ^2. 
\end{equation}
Unlike the Minkowski case, the null conformal boundary $\mathcal{I}$ of Klein space has a single connected component, which is a square Lorentzian torus fibered over a null interval. The torus metric induced from \eqref{eq:flatKlein} is \cite{Mason:2005qu}
\begin{equation}
ds^2_{\mathcal{I}} = - d\psi^2 + d\phi^2.
\end{equation}
In flat Klein space, the ``Lorentz group" is $ {SL(2,\mathbb{R})_L\times SL(2,\mathbb{R})_R }$. We refer readers to \cite{atanasov2021_22Scat} and Appendix \ref{app:torus} for additional details and discussion of flat Klein space and the celestial torus. In Klein space, null momenta can be parameterized by a positive frequency $\omega$ and a point on the momentum-space celestial torus $(\psih,\phih)$.  (Momentum-space coordinates will be denoted using hats throughout this paper.) In $(T,X,Y,Z)$ coordinates, null momenta can be parameterized explicitly as 
\begin{equation} \label{eq:qveccartesian}
    q^\mu= \omega\hat{q}^\mu(\psih, \phih) = \omega\left( \sin\psih, \cos\phih, \sin\phih, \cos\psih  \right).
\end{equation}
Note that the spacetime reversal transformation $PT: \hat{q}^{\mu} \mapsto - \hat{q}^{\mu}$ is an element of the Kleinian ``Lorentz group" and is continuously connected to the identity, in accord with $\mathcal{I}$ having only one connected component.

On Klein space, there are two different types  of  primary solutions of the massless wave equation involving different choices of raising operators within $SL(2, \mathbb{R})_L \times SL(2, \mathbb{R})_R$\cite{atanasov2021_22Scat}. The most familiar are local ``$H$-primaries'' that correspond to boost eigenstate scattering states. They are dual to local boundary operator insertions on the celestial  torus and may be constructed as Mellin transforms of plane waves.  The second type are global ``$L$-primaries.'' These  correspond to global quantum states circulating around  the boundary torus, and are related by integral transforms to the $H$-primaries. The transforms (see below) implement the  boundary state-operator correspondence.  The two types are analogs of the local  and global primaries familiar in AdS \cite{Maldacena:1998bw}.  This is discussed in detail in \cite{atanasov2021_22Scat} and here  for completeness in  Appendix \ref{app:torus}.

Analytically continuing expressions in \cite{pasterski2017conformal, pasterski2017flat}, an $H$-primary of conformal weight $\Delta$ and zero spin is given by\footnote{Note that the normalization of our $H$-primary wavefunctions differs from that of \cite{atanasov2021_22Scat} by a factor of $i^\Delta$. Also note that the point $\hat{q}\cdot X = 0$ requires a $\pm i\epsilon$ regulator prescription. Since the sign can be changed by the Kleinian rotation $ \hat{q}^{\mu} \mapsto - \hat{q}^{\mu}$, the  difference between the two choices is a simple basis change and we suppress the regulator for brevity.}
\begin{equation}\label{eq:hprimdefn}
    \varphi_{\Delta} (X; \psih, \phih)  = \frac{\Gamma(\Delta)}{(-\hat{q}(\psih, \phih)\cdot X)^\Delta},
\end{equation}
where 
\begin{equation} \label{eq:qdotX}
\hat{q}(\psih, \phih)\cdot X =   r\cos(\phih - \phi) - q\cos(\psih - \psi). 
\end{equation}
The ``operator-state" map from $H$-primaries $\varphi_{\Delta} (X; \psih, \phih)$ to $L$-primaries and descendants $\Phi^\Delta_{m,n}(X)$  is simply a Fourier transform  over the celestial torus $T^2$ :
\begin{equation}\label{eq:Lprimdefn-integralofhprim}
    \Phi^\Delta_{m,n}(X) = \frac{1}{(2\pi)^2} \int_{T^2}d\psih d\phih e^{-i(m+n)\psih - i(m-n)\phih} \varphi_{\Delta}(X; \psih, \phih).
\end{equation}
where $\Delta \in \mathbb{Z}$ and $m,n \in \mathbb{Z} - \frac{\Delta}{2}$. As discussed in Appendix \ref{app:torus}, for every $\Delta \in \mathbb{Z}$ the highest-weight primary solution has $m = n = -\frac{\Delta}{2}$ and obeys $L_1 \Phi^{\Delta}_{-\frac{\Delta}{2},-\frac{\Delta}{2}} = \bar{L}_1 \Phi^{\Delta}_{-\frac{\Delta}{2},-\frac{\Delta}{2}} = 0$. In our conventions, its explicit form is given by 
\begin{equation} \label{eq:Lhighestweight}
\Phi^{\Delta}_{-\frac{\Delta}{2},-\frac{\Delta}{2}}(X) = 2^{\Delta}\Gamma(\Delta) e^{i \psi \Delta} q^{-\Delta} . 
\end{equation}
Descendants are obtained by acting with $L_{-1},\bar{L}_{-1}$. Similarly, the lowest-weight primary $\Phi$ obeys $L_{-1} \Phi = \bar{L}_{-1} \Phi = 0$, while mixed primaries obey one of the set of conditions $L_{\pm 1} \Phi = L_{\mp 1} \Phi = 0$.

\subsection{Taub-NUT}
In this section we  derive the  expansion of the linearized spinless self-dual Taub-NUT metric into $L$-primary wavefunctions. The Kleinian Taub-NUT metric is obtained through analytic continuation of the Lorentzian Taub-NUT metric \cite{taub1951empty, nut1963empty, miller1973global}, which is a stationary, axisymmetric solution of Einstein's equations. The axisymmetry is part of a rotational $SL(2,\mathbb{R})$ symmetry that fixes the stationary direction; see Appendix \ref{app:KTN} for details. It is characterized by a mass $M$ and NUT charge $N$. In Kleinian signature, the metric is self-dual when $M=N$.  
As exploited in \cite{Crawley:2021auj,guevara2021reconstructing}, for self-dual solutions ($h_{\mu\nu}=h^+_{\mu\nu}$) with stationary direction $u^{\mu}$, the full linearized curvature is completely fixed in terms of  
\begin{equation}
    h(X) \equiv u^{\mu}u^{\nu} \kappa \ h^+_{\mu\nu}(X).
\end{equation}
This is because the only independent components of the self-dual Weyl tensor $C^+_{\mu\alpha\nu\beta}$ are
\begin{equation}
u^{\mu}u^{\nu}C^+_{\mu \alpha \nu \beta }(X)=\frac{1}{2}\nabla_\alpha \nabla_\beta \, h(X)
\end{equation}
where the derivatives are taken with respect to the flat metric in \eqref{eq:dx22}. Specializing to the rectangular coordinates \eqref{eq:cartcoord} with $u^{\mu} = (1,0,0,0)$, the field $h(X) = \kappa h^+_{00}$ is just the gravitational potential. Since this is a scalar under stationary coordinate transformations, we further introduce an expansion of the potential in a basis of scalar $L$-primaries and their descendants as\footnote{As shown in \cite{atanasov2021_22Scat}, the $L$-primaries with $\ell\in \mathbb{Z}$ are a basis for solutions of the massless scalar wave equation. (See  \cite{Freidel:2022skz,Cotler:2023qwh} for related analyses of $H$-primaries.) Assuming potentials with Coulombic falloff restricts us to $\ell>0$, which were identified as infinite-dimensional unitary representations of $SL(2,\mathbb{R})_L\times SL(2,\mathbb{R})_R$.} 
\begin{equation}
       h(X) = \frac{\kappa}{16 \pi}\sum_{\ell=1}^\infty \sum_{m,n}  h^{2-\ell}_{-m,-n}\Phi^\ell_{m,n}(X). \label{eq:TNPotentialLexpansion}
\end{equation} 
We label this according to the conformal weights of $\Phi^\ell_{m,n}$ and $h^{2-\ell}_{-m,-n}$, which become modes of wavefunctions/operators of dimension $\Delta=\ell$ and $\Delta=2-\ell$ respectively, as we will see later. 

Let us now focus on the (linearized) self-dual Taub-NUT metric. The explicit potential function is
\cite{Crawley:2021auj}\footnote{This is the analytic continuation of the standard Lorentzian potential $\kappa h_{00} = \frac{2 G_N M}{R} = \frac{2 G_N M}{\sqrt{X^2 + Y^2 + Z^2}}$  using the rotation $\theta \to i \theta$, $R \to -iR$, $M \to -i M$ of the black hole metric. The gravitational potential $\Phi_{\rm grav}$ is conventionally defined as $\Phi_{\rm grav} \equiv - \frac{\kappa}{2}h_{00}$.} 
\begin{equation}\label{eq:hx}
\begin{aligned} 
    h^{{\rm TN}}(X)  &= \frac{2M G_N}{\sqrt{Z^2 - (X^2+Y^2)}}    &= \frac{2M G_N}{\sqrt{q^2\cos^2 \psi - r^2}},
\end{aligned}
\end{equation}
defined in the region of real, positive radial coordinate termed the Kleinian Rindler wedge \cite{Crawley:2021auj}
\begin{equation} \label{eq:RindlerWedge}
    Z > \sqrt{X^2 + Y^2} \geq 0.
\end{equation}
See Appendix \ref{app:KTN} for additional discussion. 

It is very convenient to write the gravitational potential \eqref{eq:hx} as an integral over a stationary $\psih=0$ slice on the celestial torus 
\begin{equation}\label{eq:master}
    h^{{\rm TN}} (X) = \frac{M G_N}{\pi}\int_{T^2} d\phih d\psih \frac{\delta(\psih)}{ (- \hat{q}(\psih, \phih)\cdot X) },
\end{equation}
where $\hat{q}\cdot X$ is as given in \eqref{eq:qdotX}.
This form will allow us to make contact with scattering amplitudes in Section \ref{sec:hprimary-amps-wilson}. 
To connect \eqref{eq:master} to $H$-primaries on the torus, note that it is written (up to normalization) as an integral of a $\Delta=1$ $H$-primary wavefunction \eqref{eq:hprimdefn}, consistent with the observation that only the $\Delta=1$ mode will contribute due to the behavior of \eqref{eq:hx} under scaling.  By mode-expanding the delta function 
\begin{equation}
    \delta(\psih)= \frac{1}{2\pi}\sum_{\alpha\in \mathbb{Z}} e^{-i\alpha\psih},
\end{equation}
inserting it into \eqref{eq:master}, and noting that terms with even $\alpha$ vanish under the integral, we can write the $L$-primary expansion
\begin{equation}\label{eq:TNpotential-Lprimmodes}
    h^{{\rm TN}} (X) = \frac{M G_N}{2\pi^2} \sum_m 
    \int_{T^2} d\phih d\psih \frac{e^{-2im\psih}}{ q\cos (\psih-\psi) - r\cos(\phih -\phi) } =   2 M G_N  \sum_{m} \Phi^1_{m,m} (X) ,
\end{equation}
where $m \in \mathbb{Z}-\frac{1}{2}$ and in the last equality we have used the integral definition  \eqref{eq:Lprimdefn-integralofhprim}. The condition $m=n$ follows from axisymmetry ($\phi$-independence) of  the gravitational potential \eqref{eq:hx}. The infinite sum over $m$ is required by time-translation invariance.\footnote{Using \eqref{eq:timederivPhimn} it follows for the time derivative that $\partial_0  \sum_{m} \Phi^1_{m,m} \propto \sum_m \left(\Phi^2_{m+\frac{1}{2},m+\frac{1}{2}} - \Phi^2_{m-\frac{1}{2},m-\frac{1}{2}}\right)=0$. A similar calculation can be done for the rotational $SL(2,\mathbb{R})$ symmetry.} By comparison to \eqref{eq:TNPotentialLexpansion}, we obtain the simple result 
\begin{equation} \label{eq:h1mnmodes}
     h^1_{m,n}  =  \kappa \, M \, \delta_{m-n}.
\end{equation}
 Finally, we note that in usual Minkowski space the radial gravitational potential is a singlet under the rotation group. Crucially, here in \eqref{eq:TNpotential-Lprimmodes} it is instead decomposed into an infinite tower of boost eigenstates, which are simply scattering states in Klein space. As explained in \cite{Crawley:2021auj} the potential is non-singular in the Rindler wedge \eqref{eq:RindlerWedge}, which avoids the singularity at zero radius, namely at $Z=\sqrt{X^2+Y^2}$. For an expansion in small $r$, this singularity corresponds to delta-function sources at $q=0$ in the $L$-primary equation of motion \cite{atanasov2021_22Scat} (see e.g. \eqref{eq:Lhighestweight}). 

\subsection{Kerr Taub-NUT}
As observed in \cite{Crawley:2021auj}, in Kleinian signature the self-dual Kerr-Taub-NUT black hole is diffeomorphic to the self-dual Taub-NUT solution. This is a large diffeomorphism in the sense that, as we will see in Section \ref{sec:wcharges}, it endows our solution with a tower of ${\rm w}_{1+\infty}$ charges. These charges are defined intrinsically on the torus in $(2,2)$ signature, and analytically continue to asymptotic charges in Lorentzian signature \cite{Freidel:2021ytz}.

Again specializing to the potential written in rectangular coordinates \eqref{eq:hx}, the diffeomorphism becomes a shift $X^\mu \to X^\mu + a^\mu$ where $a^\mu$ is the intrinsic angular momentum. For spin $a$ in the $Z$-direction,
\begin{equation}
    a^\mu = (0,0,0,- a),
\end{equation}
where the sign accounts for the signature of the metric, we obtain the Kerr-Taub-NUT gravitational potential
\begin{equation} \label{eq:hxKerr}
  \begin{aligned}
    h^{{\rm KTN}}(X) \equiv \kappa \, h_{00}^{\textrm{KTN}}(X) = \frac{ 2 M G_N }{\sqrt{(Z-a)^2 - (X^2 + Y^2)}}, 
  \end{aligned}
\end{equation}
which is defined in the Rindler wedge $(Z-a) > \sqrt{X^2 + Y^2}\geq 0$. This also generates a linear shift in the denominator of \eqref{eq:master}:
\begin{align}\label{eq:kerrH}
        h^{{\rm KTN}}(X) &= \frac{M G_N}{\pi}\int_{T^2} d\phih d\psih \frac{\delta(\psih)}{ (-\hat{q}(\psih, \phih)\cdot (X + a))}  \\
        &= \frac{M G_N}{\pi} \sum_{\ell=0}^\infty a^\ell \int_{T^2} d\phih d\psih \frac{\delta(\psih)}{ (-\hat{q}(\psih, \phih)\cdot X )^{\ell+1}},\label{eq:kerrH2}
\end{align}
where in the second line we have expanded in $\hat{q}\cdot a$, and noted that for $a^\mu$ in the $Z$ direction, $\hat{q}\cdot a = a$ on the support of the delta function, and so can be pulled out of the integral over the torus.\footnote{Note that the spin can always be aligned with the $Z$ direction by a coordinate rotation; if the spin is not explicitly aligned with $Z$, a factor $\hat{q}(\hat{\psi},\hat{\phi})\cdot a$ will remain in the integrand and $h^{{\rm KTN}}(X)$ will be expressed with an additional convolution of modes of $\hat{q}(\hat{\psi},\hat{\phi})\cdot a$.} This is the Kleinian analog of the Lorentzian spherical harmonic expansion \eqref{eq:Lorex}. Up to normalization, the coefficient of each $a^\ell$ is an integral of an $H$-primary of weight $\Delta=\ell+1$, and their integral over the celestial torus extracts the corresponding $L$-primaries:
\begin{align}\label{eq:KTNpotential-Lprimmodes}
\begin{split}
    \frac{1}{2\pi} \int_{T^2} d\phih d\psih \frac{\delta(\psih)}{ (-\hat{q}(\psih, \phih)\cdot X )^{\ell+1}} &= \ \sum_m \frac{1}{(2\pi)^2} \int_{T^2} d\phih d\psih \frac{e^{-2im\psih}}{ ( q\cos (\psih-\psi) - r\cos(\phih -\phi))^{\ell+1}} \\
    &=  \sum_{m \in \mathbb{Z} - \frac{\ell}{2}} \frac{1}{\ell!}\Phi^{\ell+1}_{m,m} (X)
\end{split}
\end{align}
where as in the non-spinning case we have mode-expanded the delta function, used that only terms with $m \in \mathbb{Z} - \frac{\ell}{2}$ contribute, and again used the $L$-primary definition \eqref{eq:Lprimdefn-integralofhprim}. So, the Kerr-Taub-NUT potential \eqref{eq:kerrH2} takes the final form
\begin{align}
    h^{{\rm KTN}}(X) &= 2 M G_N \sum_{\ell=0}^\infty  \, \sum_{m} \, \frac{a^\ell}{\ell!}\Phi^{\ell+1}_{m,m} (X), \label{eq:kerrH3}
\end{align}
where $m \in \mathbb{Z} - \frac{\ell}{2}$. Hence, comparing to the self-dual Kerr-Taub-NUT metric mode expanded in $L$-primaries \eqref{eq:TNPotentialLexpansion},
we conclude 
\begin{equation}\label{eq:hKTNmn}
     h^{2-\ell}_{m,n}  = \kappa \, M \, \delta_{m-n} \frac{a^{\ell-1}  }{(\ell-1)!}.
\end{equation}

\section{2D Quantum State for Self-Dual Kerr Taub-NUT} \label{sec:quantum}
In this section, we construct 2D quantum states that reproduce the classical self-dual linearized (Kerr-)Taub-NUT metrics from an expectation value of positive-helicity soft graviton operators. In Section \ref{sec:wcharges} we will see that these states are moreover quantum eigenstates at the order to which we work of the global ${\rm w}_{1+\infty}$ charges characterizing the black hole solution. 

We recall from \eqref{eq:hx} that $h(X) \equiv u^{\mu}u^{\nu} \kappa h^+_{\mu\nu}(X)$ is a self-dual metric and hence described by  positive-helicity graviton operators  \cite{Penrose:1976jq}. We present the non-spinning self-dual Taub-NUT case first for clarity, then generalize to the spinning case. We begin by defining states and operators on the celestial torus, introducing the 2D vacuum state and the Goldstone operators that pair with the soft graviton operators. We then construct a canonical black hole state such that the expectation values of soft graviton operators reproduce the classical metric as well as its tower of ${\rm w}_{1+\infty}$ charges.

The expansion of a self-dual metric into conformal primary positive-helicity graviton operators involves a pairing between conformal primary wavefunctions of weight $\Delta$ and positive-helicity graviton operators of weight $2-\Delta$. This pairing is connected to a more familiar momentum-space perspective in Section \ref{sec:Wilson}.  Upon quantization, the potential $h(X)$ is promoted to an operator $\hat{h}(X)$. Our goal is to reinterpret the classical mode coefficients $h^{2-\ell}_{m,n}$ of $\Phi^{\ell}_{m,n}$ with weight $\ell$ as the 2D expectation value of the modes of a soft graviton operator $H^{2-\ell}_{m,n}$  with weight $2-\ell$. The operator $H^{2-\ell}_{m,n}$ is defined as the operator coefficients of the weight $\ell$ primary in the metric operator expansion\footnote{This is related to the conventional definition of soft graviton operators in terms of $H$-primaries in Section \ref{sec:hprimary-amps-wilson}.} 
\begin{equation}
       \hat{h}(X) \equiv \frac{\kappa}{16\pi}\sum_{\ell=1}^\infty \sum_{m,n}  H^{2-\ell}_{-m,-n} \Phi^\ell_{m,n}(X). \label{eq:potentialoperator}
\end{equation} 
In other words, we seek a 2D quantum state $|\psi\rangle $, corresponding to the black hole, for which $h(X) {=} {}\langle\psi| \hat{h}(X) |\psi\rangle $, or equivalently 
\begin{equation}
\langle\psi| H^{2-\ell}_{m,n} |\psi\rangle = h^{2-\ell}_{m,n}\
\end{equation}
where for the general self-dual Kerr-Taub-NUT solution $h^{2-\ell}_{m,n}$ is given in \eqref{eq:hKTNmn}.

\subsection{ Taub-NUT}
\label{sec:QuantumTN}
In this subsection, we reproduce the mode expansion of the Kleinian self-dual Taub-NUT metric \eqref{eq:h1mnmodes} from expectation values of soft graviton operators on the celestial torus.  Because this metric only involves the modes $\Phi^{1}_{m,n}$, in this section we consider  only $\Delta=1$ modes of the  graviton operator. Higher $\Delta$ modes will be needed below in the spinning case.

\subsubsection{$\Delta=1$  Vacuum and Goldstone Modes}
\label{sec:vacuumandGoldstones}

In order to define states on the torus, we first need to define a 2D vacuum. We will work with the weight $\left(\frac{3}{2},-\frac{1}{2}\right)$ soft graviton operator $H^{1}$, whose modes  $H^{1}_{m,n}$ are defined in \eqref{eq:potentialoperator} as the operator coefficients of $L$-primaries. Inside the wedge $|n|\le \frac{1}{2} $ the sum 
\be H^1_{\pm \frac{1}{2}}(\hat \psi+\hat \phi) = \sum _m H^1_{m,\pm \frac{1}{2}}e^{im(\hat \psi+\hat \phi)} 
\ee 
is the chiral soft current that generates supertranslations,\footnote{More precisely here they generate the self-dual combination of translations and dual supertranslations \cite{Huang:2019cja,Kol:2019nkc,Godazgar:2018dvh,Godazgar:2018qpq}.} where the components with $m= \pm \frac{1}{2}$ are global translations. For our particular massive state, we will see that in addition there is an infinite tower of non-zero soft modes with $|n|>\frac{1}{2} $ outside the wedge range that acquire nonzero expectation values in the state. 

As is conventional for holomorphic currents of weight $h=\frac{3}{2}$, we define a 2D vacuum by
\begin{equation} \label{eq:H1vac_ket}
H^1_{m,n}|0\rangle = 0, \ \ \ m > -\frac{3}{2} \ .
\end{equation}
In the usual CFT$_2$ adjoint employed here, $(H^{1}_{m,n})^\dagger=H^1_{-m,-n}$ (see  \cite{Crawley:2021_stateop} for a discussion in the celestial context) and according to BPZ conjugation \begin{equation} \label{eq:H1vac_bra}
\langle 0 | H^1_{m,n} = 0, \ \ \ m < \frac{3}{2} \, . 
\end{equation}
Note that the modes $H^{1}_{\pm \frac{1}{2},\pm \frac{1}{2}}$ annihilate the bra and ket vacua and so are indeed global symmetries. 

We will also consider the weight $\left(-\frac{1}{2}, \frac{3}{2}\right)$ Goldstone operator $G^{1}$, 
which is symplectically paired with the soft current $H^1$.  These operators have been constructed in a variety of ways in different formalisms, including from logarithmic branches in the solution space  \cite{Donnay:2018neh}, as Wilson lines \cite{Himwich:2020rro,Arkani-Hamed:2020gyp} or as edge modes at null infinity \cite{Kapec:2021eug,Nguyen:2021ydb}. Here we simply define the Goldstone modes by their 2D commutation relations with $H^1$:\footnote{The Minkowski bulk spacetime pairing is described in \cite{Donnay:2018neh} and related boundary pairings on the celestial sphere have been discussed in \cite{Himwich:2020rro,Arkani-Hamed:2020gyp,Kapec:2021eug}. We discuss a  connection to the bulk pairing in Section \ref{sec:Wilson}. } 
\begin{equation} \label{eq:H1G1pairing}
  \left[H^{1}_{m,n}, G^{1}_{m',n'}\right] = - i \delta_{m+m'} \ \delta_{n+n'} + \mathcal{O}(\kappa),
\end{equation}
where $\kappa = \sqrt{32 \pi G_N}$ and we work to leading order in $\kappa$. This two-point pairing may receive $\mathcal{O}(\kappa)$ corrections from three-point functions, and the higher-order terms in $\kappa$ involve higher-dimension interactions. 

Acting on the vacuum, we define 
\begin{equation}\label{eq:Gvacket}
G^{1}_{m',n'}|0\rangle = 0, \ \ \ m'  > \frac{1}{2},
\end{equation}
and 
\begin{equation}
\langle 0 | G^{1}_{m',n'} = 0, \ \ \ m' < -\frac{1}{2}.
\end{equation} 
 Of particular interest are the modes $-\frac{1}{2} \leq m', n' \leq \frac{1}{2}$, which do not annihilate the vacuum and are Goldstone bosons for  global translations. In  Klein space, NUT charge/momentum and ordinary energy/momentum are on equal footing, and self-dual excitations must contain equal amounts of both.  According to \eqref{eq:H1G1pairing} such excitations  are obtained by acting with exponentials of $G^1_{\pm\half,\pm \half}$. 
In a system with both mass and NUT charge present, both must be quantized; see  \cite{bunster2006monopoles, argurio2009supersymmetry, Argurio_2009} for the gravitational setup and \cite{Emond:2021lfy} for a modern perspective based on the double copy.
This implies that the modes $G^1_{\pm \frac{1}{2},\pm \frac{1}{2}}$ are  periodically identified and so not good operators on their own. They may only appear  exponentiated with quantized coefficients, as in the next section.

\subsubsection{Metric from Quantum State} \label{sec:QuantumTNState}

We now define the 2D  state $|M\rangle$ that gives the spinless Kleinian self-dual Taub-NUT metric as the expectation value of graviton operators as 
\begin{equation}\label{eq:stateTN}
|M\rangle \equiv \exp\left[i \kappa \,  M  \sum_{j}G^{1}_{j,j}\right]|0\rangle,
\end{equation}
which is a coherent state of Goldstone operators. Note that at linearized order, by \eqref{eq:Gvacket} only the terms with $j \leq \frac{1}{2}$ contribute. The sum includes the global modes of $G^{1}$ in $-\frac{1}{2} \leq j \leq \frac{1}{2}$, which carry global translation charge. 
In Section \ref{sec:Wilson} we connect this state with a Wilson line involving an integral over a time slice of the celestial torus.

The expectation value of $H^1_{m,n}$ in this state is computed from commuting $H^{1}_{m,n}$ to the right for $m \geq - \frac{1}{2}$ and to the left for $m \leq \frac{1}{2}$, and using \cref{eq:H1vac_ket,eq:H1vac_bra,eq:H1G1pairing}. 
For instance, for $m \geq -\frac{1}{2}$, we have
\begin{align}\label{eq:H1mnlinorder}
    H^{1}_{m,n} |M \rangle &=  \kappa \, M \delta_{m-n} | M  \rangle + \cdots
\end{align}
where the dots denote contributions from the $\mathcal{O}(\kappa)$ and higher terms in the commutator \eqref{eq:H1G1pairing}. There is an analogous calculation for $m \leq -\frac{1}{2}$. Comparing to \eqref{eq:h1mnmodes}, we note that  
\begin{equation}\label{eq:hmnasEV}
     \langle  M  | H^{1}_{m,n} | M \rangle =  h^1_{m,n} =    \kappa \, M \, \delta_{m-n},
\end{equation}
so the linearized Taub-NUT metric can be reproduced in terms of an expectation value in $| M \rangle$:
\begin{equation}
    h^{{\rm TN}} (X) = \frac{2M G_N}{\sqrt{Z^2 - (X^2+Y^2)}} = \frac{\kappa}{16\pi} \sum_{m,n} \langle  M  | H^{1}_{-m,-n} | M \rangle \Phi^1_{m,n}  (X). \label{eq:TNPotentialLexpansion_quantum}
\end{equation}
In general a quantum state is not uniquely fixed by its expectation values and we do not know to what extent \eqref{eq:stateTN} is unique even at the linearized level.  However the state is highly constrained by several considerations. As discussed above, the $G^{1}_{\pm \frac{1}{2}, \pm \frac{1}{2}}$ Goldstone bosons of global translation symmetry must be periodically identified and therefore must appear exponentiated.  Moreover we will see in the next section that \eqref{eq:stateTN} is an eigenstate of an infinite number of w$_{1+\infty}$ charges to the order at which we work. It is therefore  plausible that, even in the exact theory with appropriate definitions for subleading terms in $\kappa$, the exact state is constrained to take a form very similar to \eqref{eq:stateTN}.\footnote{A  mixed state  may be encountered  in some contexts.}  Interestingly, exponentiation demands that this state includes all orders in $\kappa$. We further note two pertinent  additional facts: first, that \eqref{eq:TNPotentialLexpansion_quantum} can be transformed to the fully nonlinear potential by the coordinate transformation \eqref{eq:nonlineartrans}; and second, that higher-order terms in $\kappa$ that contribute to \eqref{eq:hmnasEV} involve higher-point vacuum correlation functions of the global modes of  $G^{1}$, which can be consistently set to zero as they are negative-helicity operators that are non-interacting in a self-dual theory.

\subsection{Kerr-Taub-NUT} 
\label{sec:Kerr}
In this section, we will extend the previous result to the spinning case, recasting the $L$-primary coefficients \eqref{eq:hKTNmn} as the 2D expectation value of soft operators in the torus.  When the spin $a$ is nonzero, all the soft operators $H^k$ with  $k = 1, 0, -1, \ldots$ are involved, so we must first define the vacuum and Goldstone modes for general $k$. 

\subsubsection{2D Vacuum State and Goldstone Modes}
The general soft graviton operators $H^k$ are defined for $k = 2,1, 0, -1, \ldots$ by \eqref{eq:potentialoperator}. As in the previous section, $H^2$ does not come into play,\footnote{These are in the ideal of the soft algebra and may be consistently set to zero, but it would be of interest to find a context in which they play a role.} $H^1$ is the  leading soft graviton operator, $H^0$ is the  subleading soft graviton operator, and so on. The $H^k$ are primaries of weight $\left(\frac{k+2}{2},\frac{k-2}{2}\right)$ and each have wedge modes  $\frac{k-2}{2} \leq n \leq \frac{2-k}{2}$ \cite{guevara2021holographic}. 
In line with \eqref{eq:H1vac_ket} and \eqref{eq:H1vac_bra}, the ket vacuum is defined as standard for a set of weight $h = \frac{k+2}{2}$ currents:\footnote{This definition incorporates translation and $SL(2,\mathbb{R})_R$ invariance of the vacuum through its  invariance under the global part of the wedge modes of the leading and subleading soft graviton operators $H^{1}_{\pm \frac{1}{2}, \pm \frac{1}{2}}$ and $H^{0}_{0,0}, H^{0}_{0,\pm 1}$.  $SL(2,\mathbb{R})_L$ is an automorphism not generated by self-dual gravitons as we are only considering the self-dual sector.} 
\begin{equation} \label{eq:Hvac1}
H^k_{m,n}|0\rangle = 0, \ \ \ m > -{{k+2}\over 2},
\end{equation}
and the bra vacuum obeys 
\begin{equation} \label{eq:Hvac2}
\langle 0 | H^k_{m,n} = 0, \ \ \ m  < {{k+2}\over 2}. 
\end{equation}
Next, we introduce Goldstone operators $G^{\ell}$ that are paired with the higher soft currents. Such operators have been considered previously in \cite{Nguyen:2020hot,Donnay:2020guq, Pasterski:2021fjn, Pasterski:2021dqe, Donnay:2022sdg, Freidel:2022skz}. We consider Goldstone operators of weights $\left(\frac{\ell-2}{2},\frac{\ell+2}{2}\right)$ for $\ell = 0, 1, 2, \ldots$. We take the pairing between the modes of $H^k$ and $G^{\ell}$ to be given by
\begin{equation} \label{eq:HGpairing}
  \left[H^{k}_{m,n},G^{\ell}_{m',n'}\right] = - i \delta_{m+m'}\ \delta_{n+n'}\ \delta^{k+\ell-2} +\mathcal{O}(\kappa),
\end{equation}
where we are again working to leading order in $\kappa$. Acting on the vacuum, we define 
\begin{equation} \label{eq:Gketcondtition}
G^{\ell}_{m',n'}|0\rangle = 0, \ \ \ m' > -\frac{\ell-2}{2} \ ,
\end{equation}
and 
\begin{equation}\label{eq:Gbracondtition}
\langle 0 | G^{\ell}_{m',n'} = 0, \ \ \ m' < \frac{\ell-2}{2} \ .
\end{equation}
The definitions \eqref{eq:Hvac1}, \eqref{eq:Hvac2}, \eqref{eq:Gketcondtition}, and \eqref{eq:Gbracondtition} are consistent with the commutation relation \eqref{eq:HGpairing}. The $H^k$ and $G^{\ell}$ form an $\left[x,p\right]$ system when the commutator is nonvanishing. In such cases, one of $H^{k}_{m,n}, G^{2-k}_{-m,-n}$ annihilates both vacua, while the other does not have a well-defined action on them.  
\subsubsection{Metric from Quantum State} 
We now define the black hole state $| M,a \rangle$ that gives the Kleinian self-dual Kerr-Taub-NUT metric as the expectation value of soft graviton operators. The previous results suggest that there is a one-to-one correspondence between $L$-primaries and Goldstone modes $\Phi^{\ell}_{m,n}\longleftrightarrow G^{\ell}_{m,n}$. Then, from the metric expansion \eqref{eq:kerrH3} we infer the self-dual Kerr-Taub-NUT state to be defined by 
\begin{equation} \label{eq:BHstateKTN}
  | M,a \rangle \equiv \exp\left[i \kappa \, M 
 \ \sum_{\ell=1}^{\infty}\frac{a^{\ell-1}}{(\ell-1)!}\sum_{j}G^{\ell}_{j,j}\right]|0\rangle,
\end{equation} 
which is a coherent state of Goldstone operators.

The expectation value of $H^k_{m,n}$ in this state is computed from commuting $H^{k}_{m,n}$ to the right for $m > - \frac{k+2}{2}$ and to the left for $m < \frac{k+2}{2}$, and using \cref{eq:Hvac1,eq:Hvac2,eq:HGpairing}.\footnote{For $k\leq -2$, the modes $H^{k}_{m,n}$ with $ \frac{k+2}{2} \leq m \leq - \frac{k+2}{2}$ do not annihilate either bra or ket vacuum and can be commuted either direction. As standard, we take the vacuum one-point functions of $H^{k}_{m,n}$ to be zero in this range. } For instance, for $m > - \frac{k+2}{2}$, we use 
\begin{equation} \label{eq:Hkerrcalc}
  \begin{aligned}
    H^{k}_{m,n} | M,a \rangle &= i \kappa \, M \sum_{\ell=1}^{\infty}\frac{1}{(\ell-1)!} a^{\ell-1}\sum_{j }\left[H^{k}_{m,n},G^{\ell}_{j,j}\right]| M, a \rangle+ \cdots \\
    &= \kappa  \, M \delta_{m-n}\frac{a^{1-k}}{(1-k)!}|M,a\rangle + \cdots, \\ 
  \end{aligned}
\end{equation}
where the dots are determined by the $\mathcal{O}(\kappa)$ corrections in the commutator \eqref{eq:HGpairing}. For $m < \frac{k+2}{2}$ there is an analogous calculation for $\langle M,a | H^{k}_{m,n}$. The state $| M,a  \rangle$ is expected to transform covariantly (by shifts of $a$) under translations generated $H^{1}_{\pm \frac{1}{2}, \pm \frac{1}{2}}$ and hence cannot be an exact eigenstate, but must transform in a nontrivial representation. One expects at next order in $\kappa$ that $H^1$ mixes $G^\ell$ with  $G^{\ell+1}$ to ensure this, but all the details of the subleading terms have not been worked out. 

This reproduces the full black hole $L$-primary mode expansion by producing the classical modes \eqref{eq:hKTNmn} as an expectation value in $| M,a \rangle$: 
\begin{equation} \label{eq:Kerrmodes}
    \langle M,a | H^k_{m,n} |  M,a \rangle  = h^{k}_{m,n} = \kappa \, M \,\delta_{m-n} \frac{ a^{1-k}}{(1-k)!}
\end{equation}
This is our central result: the expansion of the linearized self-dual Kerr-Taub-NUT gravitational potential as the expectation value of soft graviton operators on the celestial torus, 
\begin{equation} \label{eq:KTNPotentialLexpansion_quantum}
    h^{{\rm KTN}}(X) = \frac{ 2 G_N M }{\sqrt{(Z+a)^2 - (X^2 + Y^2)}} = \frac{\kappa}{16\pi} \sum_{\ell=1}^\infty \sum_{m,n}  \
 \langle M,a | H^{2-\ell}_{-m,-n} | M,a\rangle \  \Phi^\ell_{m,n}(X).
\end{equation}
As in the non-spinning case, the requirements that the global modes of $G^{\ell}$ appear exponentiated as well as that the state reproduce the correct expectation values of soft graviton operators is highly constraining and suggests that the full spinning quantum black hole state resembles \eqref{eq:BHstateKTN}.

\section{${\rm w}_{1+\infty}$ Charges} \label{sec:wcharges}

A ${\rm w}_{1+\infty}$ symmetry algebra in gravity, which includes the chiral Poincar\'e algebra as a subalgebra, has been identified as an algebra of conformally soft gravitons in celestial CFT \cite{guevara2021holographic,Strominger:2021mtt}. In this section, we will connect the symmetries of the bulk Kerr-Taub-NUT metric with the $\mathrm{w}_{1+\infty}$ charges of the general state $| M,a \rangle$. Following \cite{Strominger:2021mtt,Himwich:2021dau}, the wedge modes of the soft graviton operators are related to the wedge modes of ${\rm w}_{1+\infty}$ via a light-transform: 
\begin{equation}\label{eq:wgensdefn}
{\rm w}^{p}_{m,n} = \frac{1}{\kappa} (-1)^{p+n}\Gamma(p+n)\Gamma(p-n)H^{4-2p}_{m,n},
\end{equation}
where the wedge modes of ${\rm w}^{p}_{m,n}$ have $1-p\leq n \leq p-1$. Following \cite{Strominger:2021mtt}, the left $m$ index is a current loop algebra index of a left current labelled by $p$ and $n$. Using this redefinition we find the wedge ${\rm w}_{1 +\infty}$ charges corresponding to the state $| M,a \rangle$, which follow directly from \eqref{eq:Kerrmodes}: 
\begin{equation}
\langle M,a \, | {\rm w}^p_{m,n} |  M,a\rangle  = \delta_{m-n} (-1)^{p+n}\Gamma(p+n)\Gamma(p-n)\frac{M a^{2p-3}}{(2p-3)!}, \ \ \ 1-p\leq n \leq p-1.
\end{equation}
These are just the tower of modes of higher-spin multipole moments of the Kerr-Taub-NUT solution \cite{Guevara:2018wpp}. In particular, ${\rm w}^{\frac{3}{2}}$ encodes the mass (translation charge) and ${\rm w}^2$ encodes the spin (Lorentz charge). Note that the charges beyond ${\rm w}^{\frac{3}{2}}$ are appropriately zero for the static (Taub-NUT) case.

Thus, self-dual Kleinian black holes  have an infinite tower of conserved soft charges, which are a type of soft hair \cite{Hawking:2016msc,Hawking:2016sgy}. The global charges ${\rm w}^{\frac{3}{2}}_{\pm \frac{1}{2},\pm \frac{1}{2}}$ and ${\rm w}^{2}_{0,0},{\rm w}^{2}_{0,\pm 1}$ correspond to global translation and $SL(2,\mathbb{R})$ symmetry of the self-dual vacuum, as implied by redefining \eqref{eq:Hvac1}: 
\begin{equation} 
{\rm w}^p_{m,n}|0\rangle = 0, \ \ m > p-3  \ .
\end{equation} 

\section{Connections to Scattering Amplitudes and Wilson Lines}\label{sec:hprimary-amps-wilson}

\subsection{Local $H$-Primary Basis and Three-Point Scattering Amplitudes} \label{sec:hprim-amps}
A beautiful and surprisingly simple connection between Kleinian black holes and three-point $(2,2)$-signature scattering amplitudes was recently developed in \cite{Monteiro:2020plf,Crawley:2021auj,guevara2021reconstructing,Monteiro:2021ztt}. In particular, it was shown that a self-dual classical linearized black hole metric can be reconstructed from a corresponding three-point scattering amplitude. In this section, we show that this result arises directly in our picture because boost eigenstates can be identified with scattering states in the Mellin basis. 

 In order to make more direct connection with previous literature in celestial amplitudes, we choose to work in the null coordinates\footnote{The left and right factors of the $SL(2,\mathbb{R})_L\times SL(2,\mathbb{R})_R$ symmetry of flat Klein space act independently as (real) Mobius transformations on $(z, \bar{z})$. The celestial torus is tiled by two such diamonds corresponding to incoming/outgoing particles with positive/negative frequency $\omega$ that are exchanged under the $SL(2,\mathbb{R})_L \times SL(2,\mathbb{R})_R$ spacetime-reversal transformation $PT: q^{\mu} \mapsto - q^{\mu}$, which is continuously connected to the identity in $(2,2)$ signature. }
\begin{equation} \label{eq:zcoords}
    z=\tan \frac{x^+}{2},\quad    \bar{z}=\tan \frac{x^-}{2}, \ \ \ x^{\pm} = \psih \pm \phih,
\end{equation}
adapted to a celestial diamond or plane. This  is a patch covering half of the torus; see \cite{atanasov2021_22Scat} and Appendix \ref{app:torus} for more details. On the  diamond, $(2,2)$ null vectors can be parametrized by a simple analytic continuation of the standard $(1,3)$ null vector parametrization on the celestial sphere:\begin{equation}\label{eq:qzzbar}
    q^{\mu} = \omega \hat{q}^\mu(z, \bar{z})= \omega(z+\bar{z},1+z\bar{z},z-\bar{z},1-z\bar{z}),
\end{equation}
which follows from \eqref{eq:qveccartesian} using that $\hat{q}^{\mu}$ has weight $\left(-\frac{1}{2},-\frac{1}{2}\right)$. We also use the convention
\begin{equation}
\varepsilon^{+\mu} = \partial_{z}\hat{q}^{\mu}(z,\bar{z}), \ \ \ \varepsilon^{-\mu} = \partial_{\bar{z}}\hat{q}^{\mu}(z,\bar{z}).
\end{equation}
Then, the integral expression for the Kerr Taub-NUT gravitational potential \eqref{eq:kerrH2} can be translated to an integral over a causal diamond: 
\begin{equation}
\begin{aligned}
        h^{{\rm KTN}}(X) &= \frac{M G_N}{\pi} \sum_{\ell=0}^\infty \frac{1}{\ell!}\int_{T^2} d\phih d\psih \
         \delta(\hat{q}\cdot u) \ (\hat{q}\cdot a )^\ell \ \varphi_{\ell+1} (X;\hat{\psi},\hat{\phi}) \\
        &= \frac{M G_N}{\pi} \sum_{\ell=0}^\infty \frac{1}{\ell!} \int_{\mathbb{R}^2
        } dz d\bar{z} \ (\varepsilon^- \cdot u)^2 \ (\varepsilon^+ \cdot u)^2 \ \delta(\hat{q}\cdot u) \ (\hat{q}\cdot a )^\ell \ \varphi_{\ell+1} (X;z,\bar{z}), \label{eq:hxzdiamond}
\end{aligned}   
\end{equation}
where we have used the $H$-primary definition \eqref{eq:hprimdefn} and recast \eqref{eq:kerrH2} in a covariant form. This is possible because the integral localizes to $\hat{q}\cdot u=0$, which is supported in a single diamond. 

We can now recast the potential \eqref{eq:hxzdiamond} from an $H$-primary basis with integer $\Delta$ to a principal series basis with $\Delta=1+i\lambda$,  $\lambda \in \mathbb{R}$ \cite{pasterski2017conformal,Guevara:2019ypd}. In particular, we can rewrite 
\begin{equation}
\begin{aligned}
        h^{{\rm KTN}}(X) &= \frac{M G_N}{\pi} \sum_{\ell=1}^{\infty} \frac{1}{(\ell-1)!} \int_{\mathbb{R}^2} dz d\bar{z}  \ (\varepsilon^- \cdot u)^2 \ (\varepsilon^+ \cdot u)^2 \  \delta(\hat{q}\cdot u) (\hat{q}\cdot a )^{\ell-1} \varphi_{\ell} (X; z,\bar {z}) \\
&=\frac{\kappa}{32\pi^2} u_{\mu} u_{\nu}\int_{1-i\infty}^{1+i\infty} d\Delta \int_{\mathbb{R}^2} dz d\bar{z} \ h_{\Delta} (z,\bar z)   \varphi^{\mu \nu}_{2-\Delta} (X;z,\bar {z}) \label{eq:kerrpotential-deltazzb-int1}
\end{aligned}
\end{equation}
where $ \varphi_{\Delta}^{\mu\nu}(X)=\varepsilon_{-}^{\mu}\varepsilon_{-}^{\nu} \varphi_{\Delta}(X)$ are the spinning $H$-primary wavefunctions continued from \cite{pasterski2017conformal} and 
\begin{equation} \label{eq:poshelicitygrav}
    h_{\Delta} (z,\bar z)  \equiv  \kappa \, M \ (\varepsilon^+ \cdot u)^2 \ \delta(\hat{q}\cdot u) (-\hat{q}\cdot a )^{1-\Delta}\Gamma(\Delta-1).
\end{equation}
The equivalence in \eqref{eq:kerrpotential-deltazzb-int1} follows from closing the $\Delta$ contour to the left while using the analyticity in $\Delta$ of $h_{\Delta}(z, \Bar{z})$. We now see that the integrand $h_{\Delta}(z, \Bar{z})\varphi_{2-\Delta}(X;z, \Bar{z})$  has the correct conformal dimension to render the $z,\bar{z}$ integral invariant. In the conformal $H$-primary basis, we can write this coefficient as the Mellin transform of a 4D scattering amplitude $\mathcal{M}_3(\omega,z,\bar{z})$: 
\begin{equation} \label{eq:hdeltadefn}
    h_{\Delta} (z,\bar z) = \int_0^\infty d\omega \  \omega^{\Delta-1}\ \mathcal{M}_3(\omega,z,\bar{z}) ,
\end{equation}
where 
\begin{equation} \label{eq:threepointfunc}
\mathcal{M}_3(\omega,z,\bar{z})  = \kappa \, M (\varepsilon^+ \cdot u)^2 \delta(\omega \hat{q}\cdot u)  e^{\omega \hat{q}\cdot a},
\end{equation}  
and we have used that $\delta(\omega)$ has measure zero in \eqref{eq:hdeltadefn}. 
Remarkably, this three-point amplitude is simply the graviton emission amplitude for Kerr-Taub-NUT. More precisely, it can be interpreted as a 4D three-point $\mathcal{S}$-matrix for a source with mass $M$, momentum $p^\mu=Mu^\mu$, spin vector $a^{\mu}$ and a graviton of momentum $q^\mu $ and polarization $\varepsilon^{+}_{\mu\nu}=\varepsilon^{+}_{\mu} \varepsilon^{+}_{\nu}$ \cite{Arkani-Hamed:2017jhn,Guevara:2018wpp,Huang:2019cja,Emond:2020lwi}. 

To conclude, note that by combining the above results we obtain a more familiar mode expansion 
\begin{equation}
    h^{{\rm KTN}}(X) = u^{\mu}u^{\nu}\kappa h^+_{\mu\nu} = \frac{\kappa}{32\pi^2} \int_0^{\infty} \omega d\omega \int_{\mathbb{R}^2} dz d\bar{z} 
    \ (\varepsilon^- \cdot u)^2\  \mathcal{M}_3(\omega,z,\bar{z}) \ e^{\omega \hat{q}(z, \bar{z})\cdot X}.\label{eq:kerrpotential-deltazzb-int}
\end{equation}
This casts the results of \cite{Crawley:2021auj, guevara2021reconstructing} (see also \cite{Monteiro:2020plf,Monteiro:2021ztt}) in an illuminating form, yielding the gravitational potential as an integral over three-point amplitudes. 

\subsection{Connection to Wilson Lines}
\label{sec:Wilson}

In this section, we will show that the 2D Goldstone modes $G^{\ell}$ and soft gravitons $H^k$ introduced in Section \ref{sec:quantum} can be recast in a form that is suggestive of a 4D interpretation related to the usual canonical operators of the Lorentzian theory. Here, we will simply note the connection and will leave a detailed derivation of the 2D $\leftrightarrow$ 4D map to future study. 
 
We begin by noting that the mode coefficients $H^k_{m,n}, G^\ell_{m,n}$ can be used to define operators $H^k(x^+, x^-)$ and $G^\ell (x^+, x^-)$ as in \eqref{eq:modeApp}. These can then be transformed to operators $H^{k}(z,\bar{z})$ and $G^{\ell}(z,\bar{z})$ on a celestial diamond using the coordinate transformation \eqref{eq:zcoords}. 
We can further define
\begin{align} \label{eq:apmdef}
\begin{split}
a^{+}(\omega,z,\bar{z})&\equiv\sum_{j=0}^{\infty}\omega^{j-1}H^{1-j}(z,\bar{z})\\
a^{-}(\omega,z,\bar{z})&\equiv i\sum_{\ell=0}^{\infty}\frac{(-\partial_{\omega})^{\ell}\delta(\omega)}{ \ell!}G^{\ell+1} (z,\bar{z}) ,
\end{split}
\end{align}
where  we emphasize that as defined, the operators $a^{\pm}(\omega,z,\bar{z})$ act on the 2D Hilbert space, although they are suggestive of 4D momentum-space positive and negative helicity graviton operators, respectively. Indeed, using this definition, it then follows that the 2D conformally soft gravitons, $H^k(z, \bar{z})$, are related to the 2D operators $a^+(\omega, z, \bar{z})$ by a soft limit: 
\begin{equation} \label{eq:Hlimitdef}
H^{k}(z,\bar{z}) = \lim_{\epsilon \to 0} \epsilon \int_0^\infty \frac{d\omega}{\omega}
                    ~\omega^{k+\epsilon} a^+(\omega,z,\bar{z}) = \frac{1}{(1-k)!}\lim_{\omega\to0}\partial_{\omega}^{1-k}\left(\omega a^{+}(\omega,z,\bar{z})\right).
\end{equation}
Similar relations have been found for the analogous 4D operators in \cite{Guevara:2019ypd,Pate:2019mfs,Puhm:2019zbl,Adamo:2019ipt}. It also follows that the 2D Goldstone operators are related to the 2D operators $a^-(\omega, z,\bar{z})$ 
\begin{equation}\label{eq:gfroma}
i G^{\ell}(z,\bar{z})= \int_{0}^{\infty}\frac{d\omega}{\omega}\omega^{\ell}a^{-}(\omega,z,\bar{z})\,.
\end{equation}
This is analogous to the 4D construction in \cite{Arkani-Hamed:2020gyp}.\footnote{We note that in the 4D interpretation, Goldstone operators are divergent when inserted into 4D  scattering amplitudes and may need to be regularized as $\omega \to \infty$. In addition, for $\ell=1$ there is an IR divergence as $\omega\to 0$ which the integration contour should avoid.} The commutator of $H^{k}(z,\bar{z})$ and $G^{k}(z',\bar{z}')$ is: 
\begin{equation} \label{eq:HzGzpairing}
\left[H^{k}(z,\bar{z}),G^{\ell}(z',\bar{z}')\right] = -i (2\pi)^2\delta(z-z')\delta(\bar{z}-\bar{z}')\delta^{k+\ell-2},
\end{equation}
which follows from the commutator of the modes $H^k_{m,n}, G^\ell_{m,n}$ given in \eqref{eq:HGpairing}.\footnote{Alternatively, one can use time-ordered OPEs in Lorentzian signature following \cite{Kravchuk:2018htv}.} The commutator of $a^+(\omega,z,\bar{z})$ and $a^-(\omega',z',\bar{z}')$ then follows from the definition \eqref{eq:apmdef}:
\begin{equation} \label{eq:apmcomm}
\begin{aligned} 
[a^{+}(\omega,z,\bar{z}),a^{-}(\omega',z',\bar{z}')]&= (2\pi)^2 \sum_{j=0}^{\infty}\frac{1}{j!}\omega{}^{j-1}(-\partial_{\omega'})^{j}\delta(\omega')\delta(z-z')\delta(\bar{z}-\bar{z}')\\
&=\frac{(2 \pi)^2}{\omega}\delta(\omega-\omega')\delta(z-z')\delta(\bar{z}-\bar{z}'). 
\end{aligned}
\end{equation} 
One can also check that taking the Mellin transforms \eqref{eq:Hlimitdef} and \eqref{eq:gfroma} of \eqref{eq:apmcomm} yields \eqref{eq:HzGzpairing}. Although these operators are defined to act on the 2D Hilbert space, the above relation is reminiscent of a 4D symplectic pairing, which in 4D Lorentzian signature pairs creation and annihilation operators. Its Kleinian analog would be a 4D Klein-invariant non-degenerate (i.e. symplectic) pairing. However, in the 4D Kleinian theory there is no canonical distinction between positive and negative frequencies as they can be interchanged, reflecting the fact that there is no $\mathcal{S}$-Matrix but instead an $\mathcal{S}$-vector \cite{atanasov2021_22Scat,Witten:2001kn}.  We leave the full definition and derivation of this pairing to future study.\footnote{While we set $\hbar=1$ in this work, restoring factors of $\hbar$ throughout amounts to taking $H^k\to\sqrt{\hbar}H^k$ and $G^\ell\to\sqrt{\hbar}G^\ell$, yielding an additional factor of $\hbar$ in \eqref{eq:HzGzpairing}. The relation between a 4D pairing that would give rise to 2D commutators \eqref{eq:apmcomm} and \eqref{eq:HzGzpairing} thus has the usual factor of $\hbar$ that enters in quantizing a classical canonical pairing.} 

Using the definition \eqref{eq:apmdef} of $a^+(\omega,z,\bar{z})$, we can also recast the 4D amplitude $\mathcal{M}_3(\omega,z,\bar{z})$ of the previous subsection (see \eqref{eq:threepointfunc} therein) as a 2D expectation value:
\begin{equation} \label{eq:3ptampexp}
    2\pi \mathcal{M}_3(\omega,z,\bar{z}) = \langle M, a|a^+(\omega,z,\bar{z}) |M, a \rangle,
\end{equation} 
where we again emphasize that $a^+(\omega,z,\bar{z})$ is a sum of the tower of soft graviton operators acting on the 2D Hilbert space, each of which is singled out as the coefficient of a Laurent expansion in $\omega$.
Also note that because the $G^{\ell}$ are non-interacting at the order to which we work, we have
\begin{equation}
\langle M, a|a^-(\omega,z,\bar{z}) |M, a \rangle = 0.
\end{equation}
Using the definition \eqref{eq:apmdef} of $a^-(\omega,z,\bar{z})$  also suggests a connection between the 2D (Kerr-)Taub-NUT black hole states (see  \eqref{eq:stateTN} and \eqref{eq:BHstateKTN} above) and 4D Wilson line dressings.\footnote{Connections between black holes and 4D Wilson line insertions in the $\mathcal{S}$-matrix have resurfaced recently in the context of gravitational effective field theories and the binary black hole problem, see e.g. \cite{Monteiro:2020plf,Guevara:2020xjx,Monteiro:2021ztt,Cristofoli:2021jas,Gonzo:2022tjm,DiVecchia:2022owy}.} 
To see this, consider first the zero-spin contribution from the Taub-NUT state \eqref{eq:stateTN}, containing 
\begin{align} \label{eq:Gsum}
\sum_j G^1_{j,j} & = \frac{1}{2\pi} \int_{0}^{2\pi}dx^{+}\int_{0}^{2\pi}dx^{-} \ \delta(x^{+}+x^{-}) \ G^{1}(x^{+},x^{-})\nonumber \\
& =\frac{1}{2\pi}\int dzd\bar{z} \ (\varepsilon^{+}\cdot u)^{2} \ \delta(\hat{q}(z,\bar{z})\cdot u) \ G^{1}(z,\bar{z})\nonumber \\
 & =\frac{1}{(2 \pi)^2 M}\int_0^{\infty}  d\omega \int_{\mathbb{R}^2} dzd\bar{z} \ \frac{(\varepsilon^{+}\cdot p)^{2}}{\hat{q}\cdot p+i\epsilon} \ a^{-}(\omega,z,\bar{z})+ {\rm c.c.}, 
\end{align}
where $p^{\mu} = M u^\mu$ and we have suggestively restored a factor of $(\varepsilon^{+}\cdot u)^{2}=1$ that would accompany a 4D negative-helicity graviton operator analogous to $a^-(\omega,z,\bar{z})$ (note that as defined in \eqref{eq:apmdef} $a^{-}(\omega,z,\bar{z})$ is anti-Hermitian). The sum \eqref{eq:Gsum} is suggestive of a Wilson line associated to the leading soft theorem, essentially a dressing accounting for the IR divergence as explained in \cite{Himwich:2020rro,Arkani-Hamed:2020gyp}. Indeed, we have  
\begin{equation}
\sum_j G^1_{j,j}=\int_{0}^{\infty}d\tau u^{\mu}u^{\nu}h^{-}_{\mu\nu}(x^\rho(\tau)), 
\end{equation} 
where $x^{\mu}(\tau)=u^{\mu}\tau$ is the trajectory of an eikonal 4D worldline associated to the black hole singularity and $h^{-}_{\mu\nu}$ is the 2D operator 
\begin{equation}
h^{-}_{\mu\nu}(X) =  \frac{1}{(2\pi)^2}\int_0^{\infty} \omega d\omega \int_{\mathbb{R}^2} dz d\bar{z} 
    \ \varepsilon^+_{\mu\nu} \ a^- (\omega,z,\bar{z}) \ e^{-\omega \hat{q}(z, \bar{z})\cdot X- i \epsilon  X^0} + {\rm c.c.},
\end{equation}
which is reminiscent of a quantized 4D negative-helicity mode expansion of an anti-self-dual metric. This closely resembles the 4D positive-helicity mode expansion \eqref{eq:kerrpotential-deltazzb-int}, except that it is a 2D expression and an $i\epsilon$ prescription is required for the Wilson line, which also captures the singularity as $\hat{q} \cdot X = 0$. Continuing this process, one finds that the spinning state, given by a translation of the above, corresponds to the IR finite analog for the subleading soft factors \cite{Guevara:2018wpp}, which appear exponentiated in the state $|M, a \rangle$. Explicitly, we find the 2D relation
\begin{equation}
\sum_{\ell=1}^{\infty}\sum_{j} \frac{a^{\ell-1} G^\ell_{j,j}}{(\ell-1)!} =\frac{1}{(2 \pi)^2 M}\int_0^{\infty}  d\omega \int dzd\bar{z} \ \frac{(\varepsilon^{+}\cdot p)^{2}}{\hat{q}\cdot p+i\epsilon}\exp\left(\frac{J_{\mu\nu}F^{\mu\nu}}{\varepsilon^+\cdot p}\right)a^{-}(\omega,z,\bar{z})+c.c. \ ,
\end{equation}
where $F^{\mu\nu} \equiv \varepsilon^{+[\mu}q^{\nu]}$, and $J_{\mu\nu}\equiv 2 p_{[\mu}a_{\nu]}$ is interpreted as the (dual of the) classical
angular momentum. In position space this amounts to a shift $x^{\mu}(\tau)\to x^{\mu}(\tau)-a^{\mu}$ of the 4D worldline, as expected for Kerr-Taub-NUT:  
\begin{equation} 
\sum_{\ell =1}^{\infty}\sum_{j} \frac{a^{\ell-1} G^\ell_{j,j}}{(\ell-1)!}=\int_{0}^{\infty}d\tau u^{\mu}u^{\nu} \ h^{-}_{\mu\nu}(x^\rho(\tau)-a^\rho).
\end{equation} 
Thus, in the Wilson line setting, the position-space translation of the self-dual Kerr-Taub-NUT gravitational potential \eqref{eq:hxKerr} has an analog as the translation of the 4D worldline associated to the state, as also found in \cite{Guevara:2020xjx}.

\section{Future Directions: Higher-Point Functions}

In this paper we have proposed a quantum state for self-dual Kerr-Taub-NUT with the property that 1) it is given by a coherent superposition as demanded by quantization of the charges and 2) the one-point function of the graviton operator reproduces the self-dual classical metric, as well as a tower of ${\rm w}_{1+\infty}$ charges. As explained, the double Kerr-Schild form of this metric also suggests that it could be  enough to consider its linearized form. 

Even though this provides strong evidence supporting our 2D realization of the black hole, several consistency checks should be performed to reflect the rich physics expected from self-dual black holes. The most pressing one may be the computation of higher-point ($n\geq 2$ points) correlation functions of graviton operators from our state $|M, a \rangle$. In fact, an interesting computation of conformal two-point functions in Kerr backgrounds has been presented recently in \cite{Gonzo:2022tjm}. We devote the rest of this section to showing that those results are consistent with our state.

The authors of \cite{Gonzo:2022tjm} considered the 4D two-point function of two massless scalars $\phi_{\Delta_1}(z_1,\bar{z}_1)$ and $\phi_{\Delta_2}(z_2,\bar{z}_2)$ in linearized Lorentzian Kerr and Schwarzschild backgrounds. 
We will show that we can reproduce the holomorphic OPE limit $z_{12} \to 0$ of their result (upon analytic continuation to $(2,2)$ signature) using our 2D operators. In order to connect this to the self-dual solution, we note that as in \eqref{eq:hmunu_SDplusASD} the Kleinian Kerr metric can be decomposed into a self-dual and anti-self-dual part, where the former corresponds to a positive-helicity graviton. In particular, in the limit $z_{12} \to 0$ we use the celestial OPE block \cite{guevara2021holographic,guevara2021celestial} for the 4D scalar-scalar-graviton interaction: 
\begin{equation} \label{eq:OPEblock}
  \phi_{\Delta_1}(z_1,\bar{z}_1) \phi_{\Delta_2}(z_2,\bar{z}_2) \sim  \frac{\kappa}{2} \frac{\bar{z}_{12}}{z_{12}}\int_0^1 dt \  t^{\Delta_1} (1-t)^{\Delta_2} \mathcal{G}^{+,{\rm in}}_{\Delta_1 + \Delta_2}(z_2,\bar{z}_2 + t \bar{z}_{12}),
\end{equation}
where $\mathcal{G}^{+,{\rm in}}_{\Delta}$ denotes an incoming 4D positive-helicity graviton of general dimension $\Delta$, and we have dropped the singularity in $\bar{z}_{12}$, corresponding to the other helicity graviton (which decouples in the self-dual case $M=N$). We also use the convention that the particle on the right-hand side is incoming, to account for the convention that the graviton in the black hole three-point function \eqref{eq:threepointfunc} is outgoing. By explicit calculation (see Appendix \ref{app:BHBackgroundsComparison} for details), we verify that interpreted as a 2D operator, the expectation value of the right-hand side of \eqref{eq:OPEblock} in the self-dual Kerr-Taub-NUT state $| M, a \rangle$ is consistent with the result of \cite{Gonzo:2022tjm} for the left-hand side. That is, replacing $\mathcal{G}^+_{\Delta}(z,\bar{z})$ with its 2D analog
\begin{equation} \label{eq:replacement}
 \mathcal{G}^{+}_{\Delta} (z,\bar z) \to \int_0^\infty d\omega \  \omega^{\Delta-1}\ a^+(\omega,z,\bar{z}),
\end{equation}
where $a^{+}(\omega,z,\bar{z})$ is defined as in \eqref{eq:apmdef}, we find 
\begin{equation}\label{eq:comparison}
\begin{aligned}
\langle \phi_{\Delta_1}(z_1,\bar{z}_1) &\phi_{\Delta_2}(z_2,\bar{z}_2) \rangle_{{\rm Kerr}} \\ 
&\sim \frac{\kappa}{2} \frac{\bar{z}_{12}}{z_{12}}\int_0^1 dt \  t^{\Delta_1} (1-t)^{\Delta_2} \int_0^{\infty} d\omega \ \omega^{\Delta_1 + \Delta_2 -1} \langle  M, a |a^{+}(z_2,\bar{z}_2 + t \bar{z}_{12})| M,a  \rangle,
\end{aligned}
\end{equation}
where on the RHS of \eqref{eq:comparison} we used the expression \eqref{eq:3ptampexp} for the 2D expectation value of $a^{+}(\omega,z,\bar{z})$, and on the LHS we assumed the results of \cite{Gonzo:2022tjm} (up to numerical factors related to normalization and analytic continuation, as detailed in Appendix \ref{app:BHBackgroundsComparison}). We leave it to future work to check that the full (non-linear) scalar two-point function in a Kerr Taub-NUT background can be reconstructed from our state. This may also shed light on the precise realization of ${\rm w}_{1+\infty}$ at higher points.

\section*{Acknowledgements}

We are grateful to A. Ball, U. Kol, N. Miller, R. Monteiro, S.A. Narayanan, D. O'Connell, A. Puhm and A. Sharma for useful conversations. A.G. acknowledges support from the Harvard Society of Fellows. This work was supported in part by NSF PHY-2207659 and the Black Hole Initiative at Harvard University, which is funded by grants from the John Templeton Foundation and the Gordon and Betty Moore Foundation. E.C. acknowledges support from a PGS D fellowship from the Natural Sciences and Engineering Research Council of Canada (NSERC) as well as the Ashford Fellowship at Harvard University.

\appendix

\section{Kleinian Kerr-Taub-NUT Metric}\label{app:KTN}

We work with a Kleinian Kerr-Taub-NUT metric in pseudo-spherical $(t,r,\theta,\phi)$ coordinates that are obtained by analytic continuation of $\theta \to i \theta$, $r \to - i r$, and $M \to - i M$ of the parameters in the Lorentzian Kerr-Taub-NUT metric \cite{miller1973global}. It takes the form (setting $G_N =1$) 
\begin{equation}\label{eq:KTN_metric_Klein}
\begin{aligned}
    ds^2 &= - \Sigma (\frac{d{r}^2}{\Delta} - d {\theta}^2) + \frac{\sinh^2 \theta }{\Sigma} (a dt + \rho d \phi)^2 - \frac{\Delta}{\Sigma}(dt - A d \phi)^2 \\
    \Sigma &= {r}^2 - (N + a \cosh \theta)^2 \\
    \Delta &= {r}^2 - 2 M r +  N^2 - a^2\\
    A &= -a \sinh^2 \theta - 2 N \cosh \theta \\
    \rho &= {r}^2 -  N^2 - a^2 = \Sigma - a A.
\end{aligned}
\end{equation}
Note that the line element $ds^2$ differs by signs relative to \cite{Crawley:2021auj}, consistent with a differing overall sign in the line element of flat space \eqref{eq:cartcoord}.\footnote{Our flat metric differs from the analytic continuation used in \cite{Crawley:2021auj}, where the flat Klein metric convention was 
\begin{equation*}
    ds^2 = dT^2 - dX^2 -dY^2 +dZ^2. 
\end{equation*}
The metric \ref{eq:KTN_metric_Klein} is obtained by the rotation $t \to - i t$, $a \to -ia$, $r \to - i r$, $M \to - i M$, and $N \to - iN$ of the corresponding metric in \cite{Crawley:2021auj}.}
This metric reduces to the Taub-NUT metric when $a=0$, and the Schwarzschild metric when $N=0$ and $a=0$. The Kerr-Taub-NUT metric contains two singularities at $r = \pm(N +a\cosh\theta)$, which can be seen from the Kretschmann scalar
\begin{equation}
 R_{\mu \nu \rho \sigma} R^{\mu \nu \rho \sigma} = 24 \left(\frac{(M - N)^2}{(N - r+a\cosh\theta)^6} + \frac{(M + N)^2}{(N + r+a\cosh\theta)^6}\right)\,. \label{eq_22KTNKretsch}
\end{equation}
When $M=N$, the metric is self-dual, satisfying
\begin{equation}
     R_{\mu \nu \rho \sigma} = \frac{1}{2} \varepsilon_{\mu \nu \alpha \beta} R^{\alpha \beta}_{\;\;\;\; \rho \sigma}, 
\end{equation}
and has a single singularity at $r = -M-a\cosh\theta$. 
Restoring factors of $G_N$, the full (nonlinear) gravitational potential of the self-dual Kerr-Taub-NUT metric is 
\begin{equation} \label{eq:nonlinearpot}
u^{\mu} u^{\nu}  g_{\mu\nu}(x) + 1  =  \frac{2M G_N}{r + a\cosh\theta + G_N M}.
\end{equation}
As discussed in \cite{Crawley:2021auj}, the coordinate transformations 
\begin{equation} \label{eq:nonlineartrans}
\begin{aligned}
    X + i Y & \equiv \sqrt{(r + G_N M)^2 - a^2 } e^{i \phi}\sinh\theta \\
    Z &\equiv - (r + G_N M) \cosh \theta
\end{aligned}
\end{equation}
for any spin $a \in \left[0,M\right]$ can be used to implement diffeomorphisms between self-dual Kerr-Taub-NUT solutions with varying spins. Applying this transformation to \eqref{eq:nonlinearpot} yields the potential \eqref{eq:hxKerr} in the Rindler wedge $(Z-a) > \sqrt{X^2 + Y^2}\geq 0$ of flat Klein space.\footnote{Note that $Z>0$ when $r < -2G_N M$.} After writing the metric \eqref{eq:KTN_metric_Klein} in the coordinates \eqref{eq:nonlineartrans}, the Kerr-Taub-NUT and Taub-NUT ($a=0$) metrics are related by the diffeomorphism
\begin{equation}
    Z\to Z+a.
\end{equation}
At linearized order in $G_N$, the self-dual Kerr-Taub-NUT gravitational potential takes the form
\begin{equation}
  \begin{aligned}
u^{\mu} u^{\nu}  g_{\mu\nu}(x) + 1 =  \frac{2M G_N}{r + a\cosh\theta} + \mathcal{O}(G_N^2).
  \end{aligned}
\end{equation}
This can be transformed to the nonlinear potential \eqref{eq:nonlinearpot} via a Talbot shift $r \to r + G_N M$ \cite{Talbot1969}. The same statement applies for the curvature invariants, see e.g. \cite{Crawley:2021auj}. So, the coordinate transformation \eqref{eq:nonlineartrans} can be used to relate the linearized and non-linear gravitational potentials.

The symmetries of the self-dual Taub-NUT metric ($a = 0$) are $U(1) \times SL(2,\mathbb{R})$, where $U(1)$ is the (compactified) time translation symmetry and the $SL(2,\mathbb{R})$ is the diagonal subgroup of the $SL(2,\mathbb{R})_L \times SL(2,\mathbb{R})_R$ symmetry of flat Klein space corresponding to the little group of the time direction. In the self-dual Kerr-Taub-NUT metric, this symmetry is broken by the spin direction to $U(1) \times U(1)$. See \cite{Crawley:2021auj} for explicit symmetry vector fields (note the differing signature convention).

\section{Operators on the Celestial Torus}
\label{app:torus}

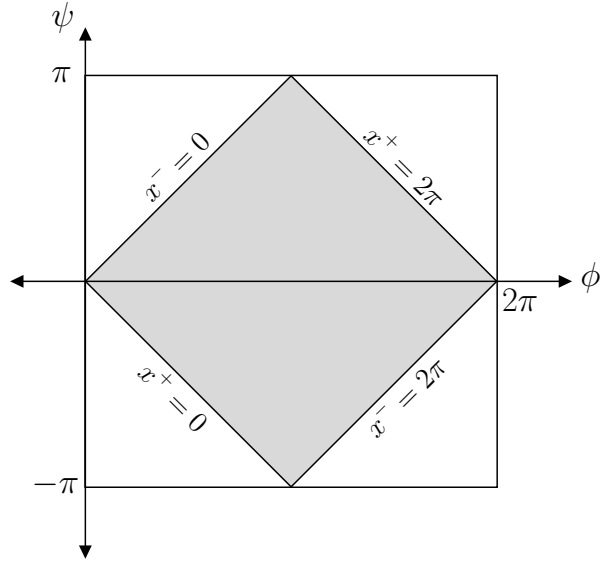
\begin{figure}
    \centering
    \resizebox{0.5\textwidth}{!}    {
        \input{diamonds.tex}
    }
\caption{The celestial torus in $\psi$ and $\phi$ coordinates, where each of $\psi$ and $\phi$ is $2\pi$-periodic. Lines of constant $x^\pm = \psi \pm \phi$ are labelled. In $x^+, x^-$ coordinates, the torus is tiled into two causal diamonds (shaded in grey and white).}
\label{fig:diamonds}
\end{figure}

The null conformal boundary $\mathcal{I}$ of Klein space is a square Lorentzian torus with metric
\begin{equation}
ds^2_{\mathcal{I}} = - d\psi^2 + d\phi^2,  \ \ \psi \sim \psi + 2 \pi, \ \ \ \phi \sim \phi + 2 \pi.
\end{equation}
This appendix follows \cite{atanasov2021_22Scat}, to which we refer readers for additional details and discussion. The ``Lorentz" symmetries of flat Klein space are 
$SO(2,2) \cong \frac{SL(2,\mathbb{R})_L\times SL(2,\mathbb{R})_R }{-1_L \times -1_R}$. The left and right factors of  $SL(2,\mathbb{R})_L\times SL(2,\mathbb{R})_R$ act separately on 
\begin{equation}
x^{\pm} = \psi \pm \phi,
\end{equation} 
which have periodicity conditions that are not independent: 
\begin{equation}
    (x^+, x^-) \sim (x^+ +2\pi, x^- + 2\pi) \sim (x^+ +2\pi, x^- - 2\pi).
\end{equation}
As illustrated by the shading in Figure \ref{fig:diamonds}, the celestial torus can be split into two celestial diamonds, defined by $x^+,x^- \in [0,2\pi)$ and its complement. 

On each celestial diamond, the $SL(2,\mathbb{R})_L\times SL(2,\mathbb{R})_R$ further acts as (real) Mobius transformations on 
\begin{equation}\label{eq:xplusminusdefnapp}
    z = \tan\frac{x^+}{2}, \ \ \  \bar{z} = \tan\frac{x^-}{2}. 
\end{equation}
While the coordinates $x^{\pm}$ cover the whole torus, $(z,\bar{z})$ are only good coordinates on one causal diamond because $\tan \frac{x^+}{2} = \tan \frac{x^+ + 2\pi}{2}$. In the $z,\bar{z}$ coordinates, switching diamonds amounts to changing the sign of $\omega$ in \eqref{eq:qzzbar}, or performing the spacetime-reversal transformation $PT: q^{\mu} \mapsto - q^{\mu}$, which is continuously connected to the identity in $SL(2,\mathbb{R})_L \times SL(2,\mathbb{R})_R$.

The mode expansion of a field/operator $\mathcal{O}^{h, \bar{h}}(x^+, x^-)$ on the celestial torus is given by 
\begin{equation}
    \label{eq:modeApp}
\mathcal{O}^{h, \bar{h}}(x^+, x^-) = \sum_{m,n} \mathcal{O}^{h,\bar{h}}_{m,n} 
\  e^{im x^+} e^{i n x^{-}} = \sum_{m,n} \mathcal{O}^{h,\bar{h}}_{m,n} \ e^{i(m+n)\psi + i(m-n)\phi},
\end{equation}
where the modes are given by integrating over a causal diamond: 
\begin{equation} \label{eq:modecommutator}
    \mathcal{O}^{h, \bar{h}}_{m,n} = \frac{1}{(2\pi)^2}  \int_0^{2\pi} dx^+ \int_0^{2\pi} dx^- \mathcal{O}^{h, \bar{h}}(x^+, x^-) \ e^{-imx^+} e^{-inx^-}.
\end{equation}
Here, spatial periodicity requires $m-n \in \mathbb{Z}$ and timelike periodicity requires $m+n \in \mathbb{Z}$. In order that the modes create the primary and descendant states associated to $\mathcal{O}^{h,\bar{h}}$ and transform properly under conformal transformations, we need $m,n \in \mathbb{Z} - h$, which then implies $h \pm \bar{h} \in \mathbb{Z}$ and therefore $h \in \frac{\mathbb{Z}}{2}$. 
The mode integral can be extended to an integral over the full torus by noting that the other diamond is given by $x^+ \to x^+ + 2\pi$ with $x^-$ fixed, under which the phases acquired by the integrand of \eqref{eq:modecommutator} vanish when $m \in \frac{\mathbb{Z}}{2}$.

As a special case, we define the global scalar $L$-primary wavefunctions as modes of the scalar $H$-primary wavefunctions \eqref{eq:hprimdefn}:
\begin{align}\label{eq:Lprimdefn-xpm}
    \Phi^\Delta_{m,n}(X) &= \frac{1}{(2\pi)^2}\int_{0}^{2\pi}d\xph \int_{0}^{2\pi} d\xmh e^{-im\xph - in\xmh} \varphi_{\Delta}(X; \xph, \xmh),
\end{align}
where $m,n \in \mathbb{Z} - \frac{\Delta}{2}$, and $\Delta\in \mathbb{Z}$, so the integral can be replaced by half the integral over the entire torus, giving the form in $(\psih, \phih)$ in \eqref{eq:Lprimdefn-integralofhprim}. 

A simple action of the left and right generators of $SL(2, \mathbb{R})_L \times SL(2, \mathbb{R})_R$ is found by making the coordinate choice for flat Klein space
\begin{equation}
z = r e^{i \phi}, \ \ \ w = q e^{i \psi},
\end{equation}
in which the metric \eqref{eq:flatKlein} becomes
\begin{equation}
ds^2 = dz d\bar{z} - d w d \bar{w}, 
\end{equation}
and the action of the symmetry is generated by the six Killing vector fields \cite{atanasov2021_22Scat}
\begin{equation} \label{eq:Lvecs}
\begin{aligned}
L_1 = \bar{z} \partial_w + \bar{w}\partial_z, \qquad &\bar{L}_1 = z \partial_w + \bar{w}\partial_{\bar{z}} \\
L_0 = \frac{1}{2}\left(z \partial_z + w \partial_w - \bar{z}\partial_{\bar{z}} - \bar{w} \partial_{\bar{w}} \right), \qquad &\bar{L}_0 = \frac{1}{2}\left(- z \partial_z + w \partial_w + \bar{z}\partial_{\bar{z}} - \bar{w} \partial_{\bar{w}} \right) \\
L_{-1} = -z \partial_{\bar{w}} - w \partial_{\bar{z}}, \qquad &\bar{L}_{-1} = -\bar{z}\partial_{\bar{w}} - w \partial_z. 
\end{aligned}
\end{equation}
They obey
\begin{equation}
\left[ L_n,L_m\right] = (n-m) L_{m+n},
\end{equation}
and similarly for the $\bar{L}_n$. 
On the boundary celestial torus, these reduce to 
\begin{equation}
L_n = - \frac{i}{2} e^{- i n (\psi + \phi)}\left(\partial_{\psi} + \partial_{\phi}\right), \ \ \ \bar{L}_n = -\frac{i}{2} e^{-in(\psi - \phi)} \left(\partial_{\psi} - \partial_{\phi}\right), \ \ \ n = -1,0,1.
\end{equation}
Certain primaries of highest/lowest/mixed weight are annihilated by various combinations of the $L_{\pm 1}$ ($\overline{L}_{\pm 1}$) of the left (right) global $SL(2, \mathbb{R})_L \times SL(2, \mathbb{R})_R$. In particular, for every $\Delta \in \mathbb{Z}$ the highest-weight primary solution is $\Phi^{\Delta}_{-\frac{\Delta}{2},-\frac{\Delta}{2}}$ and is defined to obey $L_1 \Phi^{\Delta}_{-\frac{\Delta}{2},-\frac{\Delta}{2}} = \bar{L}_1 \Phi^{\Delta}_{-\frac{\Delta}{2},-\frac{\Delta}{2}} = 0$. In our conventions, the highest-weight primary \cite{atanasov2021_22Scat} is
\begin{equation}
\Phi^{\Delta}_{-\frac{\Delta}{2},-\frac{\Delta}{2}} =  2^{\Delta} \Gamma(\Delta) \bar{w}^{-\Delta} =  2^{\Delta}\Gamma(\Delta) e^{i \psi \Delta} q^{-\Delta} . 
\end{equation}
Similarly, the lowest-weight primary $\Phi$ obeys $L_{-1} \Phi = \bar{L}_{-1} \Phi = 0$, while mixed primaries obey one of the set of conditions $L_{\pm 1} \Phi = L_{\mp 1} \Phi = 0$.  For $\Delta>0$, the highest-weight primary  and its descendants form an infinite-dimensional representation of $SL(2, \mathbb{R})_L \times SL(2, \mathbb{R})_R$. Moreover, since spacetime translations shift the conformal weight, the set of all $L$-primaries and their descendants form a representation of the Poincar\'e group.

These generators \eqref{eq:Lvecs} can be related to the $H$-generators, which are the usual $SL(2,\mathbb{R})$ operators in a basis that diagonalizes boosts towards $(\xph, \xmh)$. The explicit relation is \cite{atanasov2021_22Scat}
\begin{equation}
\begin{aligned}
H_0^{\hat{x}} = \frac{1}{2}(e^{i\xph}L_1 - e^{-i\xph}L_{-1})  , \qquad 
&\bar{H}_0^{\hat{x}} = \frac{1}{2}(e^{i\xmh}\bar{L}_1 - e^{-i\xmh}\bar{L}_{-1})  \\  
H^{\hat{x}}_{\pm 1} = iL_0 \mp \frac{i}{2} (e^{i\xph}L_1 + e^{-i\xph}L_{-1}) , \qquad &\bar{H}^{\hat{x}}_{\pm 1} = i\bar{L}_0 \mp \frac{i}{2} (e^{i\xmh}\bar{L}_1 + e^{-i\xmh}\bar{L}_{-1}).
\end{aligned}
\end{equation}
In this paper, we focus on $L$-primaries with $m=n$. For such solutions, we have 
\begin{equation}
\begin{aligned}
i \partial_0 \Phi_{m,n}^\ell (X) =& \frac{i}{(2\pi)^2}\int_{T^2} d\phih d\psih e^{-i(m+n)\psih} e^{-i(m-n)\phih} \left(\ell \hat{q}_0(\psih,\phih) \right) \frac{\Gamma(\ell)}{(-\hat{q}\cdot X)^{\ell +1}} \\
=& - \frac{i}{(2\pi)^2}\int_{T^2} d\phih d\psih e^{-i(m+n)\psih} e^{-i(m-n)\phih} \frac{(e^{i\psih} - e^{-i\psih})}{2i} \varphi_{\ell +1}(X;\psih, \phih) \\ 
=& \frac{1}{2}\left(\Phi^{\ell+1}_{m+1/2,n+1/2} (X) -  \Phi^{\ell+1}_{m-1/2,n-1/2} (X)\right). \label{eq:timederivPhimn}
\end{aligned}
\end{equation}

\section{Detailed Comparison to Celestial Scattering in Kerr Backgrounds} \label{app:BHBackgroundsComparison} 

In this appendix, we connect our results to recent work on celestial holography in Kerr-Schild backgrounds in \cite{Gonzo:2022tjm}, in which the authors computed 4D celestial two-point scalar amplitudes on various backgrounds, including Schwarzschild and Kerr.

To make this connection explicit, we begin with the 4D two-point function of two massless scalars $\phi_{\Delta_1}(z_1,\bar{z}_1)$ and $\phi_{\Delta_2}(z_2,\bar{z}_2)$ in a Kerr background, which is written as (3.16) in \cite{Gonzo:2022tjm}:
\begin{equation} \label{eq:AndreaKerr}
  \begin{aligned}
    \widetilde{\mathcal{M}}_{2,{\rm Kerr}}^{(1)}(\Delta_1,\Delta_2) &= 8 \pi^2 G_N M \frac{a^{1-\Delta_1 - \Delta_2}}{|z_{12}|^2} \left(\frac{1 + |z_1|^2}{1+|z_2|^2}\right)^{\Delta_2} \frac{(1+|z_2|^2)^{\Delta_1 + \Delta_2}}{\left[2(|z_1|^2 - |z_2|^2)\right]^{\Delta_1 + \Delta_2-1}} \\
    &\qquad \times \left[\mathcal{I}''_{\rm even}(\Delta_1 + \Delta_2 - 1) - \frac{z_1 \bar{z}_2 - z_2\bar{z}_1}{|z_1|^2 - |z_2|^2} \mathcal{I}''_{\rm odd}(\Delta_1 + \Delta_2 - 1) \right], \\
  \end{aligned}
\end{equation}
where the functions $\mathcal{I}''$ are defined by 
\begin{equation}
  \mathcal{I}''_{\rm even}(s) \equiv \alpha^s\int_0^{\infty} d \omega \ \omega^{s-1}\cosh{\alpha \omega}, \ \ \  \mathcal{I}''_{\rm odd}(s) \equiv \alpha^s\int_0^{\infty} d \omega \ \omega^{s-1}\sinh{\alpha \omega},
\end{equation}
with $\alpha = 2 a \frac{|z_1|^2 - |z_2|^2}{1 + |z_2|^2}$.
Note that the authors of \cite{Gonzo:2022tjm} work in $(1,3)$ signature with the usual conventions 
\begin{equation}
  q^{\mu}(z,\bar{z}) = \eta \omega \hat{q}^{\mu}(z,\bar{z}) = \eta \omega(1 {+} |z|^2, z{+} \bar{z},-i(z{-}\bar{z}), 1 {-} |z|^2), \ \ u^{\mu} = (1,0,0,0), \ \ a^{\mu} = (0,0,0,a).
\end{equation}
and consider a two-point function of one incoming ($\eta_1 = -1$) and one outgoing ($\eta_2 = +1$) scalar. For this appendix, we will work in the conventions of \cite{Gonzo:2022tjm} for explicit comparison. Upon rotation of the spin $a \to i a$ (this can be interpreted as part of a continuation of the $Z$ coordinate to match Kleinian signature), note that using the fact that for ${\rm Re}(\alpha) = 0$,
\begin{equation}
  \mathcal{I}''_{\rm even}(s) = i^s \cos\left(\frac{\pi s}{2}\right) \Gamma(s), \ \ \  \mathcal{I}''_{\rm odd}(s) = i^{s+1} \sin\left(\frac{\pi s}{2}\right) \Gamma(s), \ \ \ 0 < {\rm Re}(s) < 1,
\end{equation}
and analytically continuing to other values of $s$, we find (note here $a$ is the Kleinian spin)
\begin{equation} \label{eq:AndreaContinued}
  \begin{aligned}
    \widetilde{\mathcal{M}}_{2,{\rm Kerr}}^{(1)}(\Delta_1,\Delta_2) &= 8 \pi^2 G_N M \frac{a^{1-\Delta_1 - \Delta_2}}{|z_{12}|^2} \left(\frac{1 + |z_1|^2}{1+|z_2|^2}\right)^{\Delta_2} \frac{(1+|z_2|^2)^{\Delta_1 + \Delta_2}}{\left[2(|z_1|^2 - |z_2|^2)\right]^{\Delta_1 + \Delta_2-1}} i^{\Delta_1 + \Delta_2 -1} \\
    &\qquad \times  \left[\cos\left(\frac{\pi}{2}(\Delta_1 {+} \Delta_2 {-}1)\right)  - \frac{z_1 \bar{z}_2 - z_2\bar{z}_1}{|z_1|^2 - |z_2|^2}  i \sin\left(\frac{\pi}{2}(\Delta_1 {+} \Delta_2 {-}1)\right) \right] \Gamma(\Delta_1 {+} \Delta_2{-}1), \\
  \end{aligned}
\end{equation}
Now, we show that we can reproduce this result in the holomorphic OPE limit $z_{12} \to 0$ using a 2D expectation value in our black hole state $|M, a \rangle$. We found the covariant $(2,2)$ result (using \eqref{eq:Kerrmodes})
\begin{equation}
  \langle M, a | H^k(z,\bar{z})| M, a\rangle = 2 \pi \kappa \, M (\varepsilon^+ \cdot u)^2 \delta(\hat{q}\cdot u)\frac{(\hat{q} \cdot a)^{1-k}}{(1-k)!}.
\end{equation}
Defining the general 2D spin-2 operator $\mathcal{O}_{(\Delta,2)}(z,\bar{z})$ by
\begin{equation}
 \mathcal{O}_{(\Delta,2)} (z,\bar z) \equiv \int_0^\infty d\omega \  \omega^{\Delta-1}\ a^+(\omega,z,\bar{z}),
\end{equation}
where $a^{+}(\omega,z,\bar{z})$ is defined in \eqref{eq:apmdef},
we therefore have (see also \eqref{eq:poshelicitygrav})
\begin{equation}
  \langle M, a | \mathcal{O}_{(\Delta,2)}(z,\bar{z}) |M, a \rangle = 2 \pi \kappa \, M (\varepsilon^+ \cdot u)^2 \delta(\hat{q}\cdot u)(- \hat{q} \cdot a)^{1-\Delta} \Gamma(\Delta - 1). 
\end{equation}
Note that
\begin{equation}
  \langle M,a | H^k(z,\bar{z})| M, a\rangle = {\rm Res}_{\Delta = k} \ \langle M,a | \mathcal{O}_{(\Delta,2)}(z,\bar{z})|M,a\rangle, 
\end{equation}
which follows from \eqref{eq:Hlimitdef}.

Using our construction, the emission of a positive-helicity graviton from a Kerr black hole will capture its self-dual part, which we can evaluate with $M = N$ to compare to results of our paper. This positive-helicity graviton can then interact with two scalars $\phi_{\Delta_1}(z_1,\bar{z}_1),\phi_{\Delta_2}(z_2,\bar{z}_2)$ through a four-point amplitude. In the limit $z_1 \to z_2$, we use the OPE block \cite{guevara2021holographic,guevara2021celestial} for the 4D scalar-scalar-graviton interaction: 
\begin{equation}
  \phi_{\Delta_1}(z_1,\bar{z}_1) \phi_{\Delta_2}(z_2,\bar{z}_2) \sim \frac{\kappa}{2} \frac{\bar{z}_{12}}{z_{12}}\int_0^1 dt \  t^{\Delta_1} (1-t)^{\Delta_2} \mathcal{G}^{+,{\rm in}}_{\Delta_1 + \Delta_2}(z_2,\bar{z}_2 + t \bar{z}_{12}),
\end{equation}
where we use the convention that the particle on the right-hand side is incoming, to account for the fact that this particle is outgoing from the black hole.\footnote{Note that in the standard all-outgoing holomorphic OPE conventions, this is simply the usual (as in \cite{Himwich:2021dau})
\begin{equation*}
  \phi_{\Delta_1}(z_1,\bar{z}_1) \phi_{\Delta_2}(z_2,\bar{z}_2) \sim \frac{\kappa}{2} \frac{\bar{z}_{12}}{z_{12}}B(\Delta_1 + 1, \Delta_2 + 1) \mathcal{G}^{-}_{\Delta_1 + \Delta_2}(z_2,\bar{z}_2) + {\rm antiholomorphic \  descendants}.
\end{equation*}} Taking the $z_1 \to z_2$ limit of the two point function thus extracts the self-dual contribution of the background. Now let's take the expectation value of the right-hand side in our black hole state $|M, a\rangle$ using the conventions from \cite{Gonzo:2022tjm}, and replacing the 4D operator with the analogous 2D operator: 
\begin{equation}
\mathcal{G}^{+,{\rm in}}_{\Delta_1 + \Delta_2}(z_2,\bar{z}_2 + t \bar{z}_{12}) \to \mathcal{O}_{(\Delta_1 + \Delta_2,2)}(z_2,\bar{z}_2 + t \bar{z}_{12}). 
\end{equation}
Defining 
\begin{equation}
\langle \phi_{\Delta_1} \phi_{\Delta_2}\rangle_{ \rm 2D, Kerr} 
    \sim \frac{\kappa}{2} \frac{\bar{z}_{12}}{z_{12}}\int_0^1 dt \  t^{\Delta_1} (1-t)^{\Delta_2} \langle M, a | \mathcal{O}_{(\Delta_1 + \Delta_2,2)}(z_2,\bar{z}_2 + t \bar{z}_{12})| M, a \rangle,
\end{equation}
and using that in the conventions of \cite{Gonzo:2022tjm}, 
\begin{equation}
\begin{aligned}
 \langle M&, a  | \mathcal{O}_{(\Delta_1 + \Delta_2,2)}(z_2,\bar{z}_2 + t \bar{z}_{12}) |M, a \rangle \\
 &=2 \pi  \kappa \, M  (\varepsilon^-(z_2,\bar{z}_2{+}t \bar{z}_{12}) \cdot u)^2 \delta(\hat{q}(z_2,\bar{z}_2 {+} t \bar{z}_{12})\cdot u)(- \hat{q}(z_2,\bar{z}_2 {+} t \bar{z}_{12}) \cdot a)^{1-\Delta_1 - \Delta_2} \Gamma(\Delta_1 + \Delta_2 - 1), \\
 &=  2 \pi \kappa \, M  z_2^2 \ \delta(1 {+} |z_2|^2 {+} t z_2 \bar{z}_{12}) \left[-a(1 {-} |z_2|^2 {-} t z_2 \bar{z}_{12}) \right]^{1-\Delta_1 - \Delta_2} \Gamma(\Delta_1 + \Delta_2 - 1),
 \end{aligned}
\end{equation}
where we note that we use $\varepsilon^-(z_2,\bar{z}_2 + t \bar{z}_{12})$ to account for the fact that the 4D  graviton is incoming to the scalar vertex, we find 
\begin{equation}
  \begin{aligned}
    \langle \phi_{\Delta_1} \phi_{\Delta_2}&\rangle_{\rm 2D, Kerr} \\
    &\sim  (-1)^{\Delta_1} \frac{32 \pi^2 M G_N }{z_{12}\bar{z}_{12}} \  \left(1+|z_2|^2\right)^{\Delta_1}\left(1+z_2\bar{z}_1\right)^{\Delta_2} (-2 a \  z_2\bar{z}_{12})^{1 - \Delta_1 - \Delta_2} \Gamma(\Delta_1 + \Delta_2 - 1), \\ 
  \end{aligned}
\end{equation}
Now, comparing to the $z_1 \to z_2$ holomorphic limit of \eqref{eq:AndreaContinued} from \cite{Gonzo:2022tjm}, we find that
\begin{equation} 
  \begin{aligned}
    \lim_{z_1 \to z_2} \widetilde{\mathcal{M}}_{2,{\rm Kerr}}^{(1)}(\Delta_1,\Delta_2) &= (-1)^{\Delta_1}  \lim_{z_1 \to z_2} \tfrac{1}{4}\langle \phi_{\Delta_1} \phi_{\Delta_2}\rangle_{\rm 4D, Kerr},
  \end{aligned}
\end{equation}
where our normalizations differ and the factor $(-1)^{\Delta_1}$ is related to incoming/outgoing conventions.

\bibliography{main}
\bibliographystyle{utphys}

\end{document}

%% file: diamonds.tex
\tikzset{every picture/.style={line width=0.75pt}} 

\begin{tikzpicture}[x=0.75pt,y=0.75pt,yscale=-1,xscale=1]

\draw   (172,91) -- (447,91) -- (447,366) -- (172,366) -- cycle ;
\draw[fill=gray!30]  (309.5,91.27) -- (446.73,228.5) -- (309.5,365.73) -- (172.27,228.5) -- cycle ;
\draw    (172.2,410.8) -- (172.2,61.8) ;
\draw [shift={(172.2,58.8)}, rotate = 90] [fill={rgb, 255:red, 0; green, 0; blue, 0 }  ][line width=0.08]  [draw opacity=0] (8.93,-4.29) -- (0,0) -- (8.93,4.29) -- cycle    ;
\draw [shift={(172.2,413.8)}, rotate = 270] [fill={rgb, 255:red, 0; green, 0; blue, 0 }  ][line width=0.08]  [draw opacity=0] (8.93,-4.29) -- (0,0) -- (8.93,4.29) -- cycle    ;
\draw    (125,228.5) -- (494,228.5) ;
\draw [shift={(497,228.5)}, rotate = 180] [fill={rgb, 255:red, 0; green, 0; blue, 0 }  ][line width=0.08]  [draw opacity=0] (8.93,-4.29) -- (0,0) -- (8.93,4.29) -- cycle    ;
\draw [shift={(122,228.5)}, rotate = 0] [fill={rgb, 255:red, 0; green, 0; blue, 0 }  ][line width=0.08]  [draw opacity=0] (8.93,-4.29) -- (0,0) -- (8.93,4.29) -- cycle    ;

\draw (148,40.4) node [anchor=north west][inner sep=0.75pt]  [font=\LARGE]  {${\psi }$};
\draw (501,213.9) node [anchor=north west][inner sep=0.75pt]  [font=\LARGE]  {${\phi }$};
\draw (148,84) node [anchor=north west][inner sep=0.75pt]  [font=\LARGE]  {$\pi $};
\draw (388.95,152.66) node  [font=\Large,rotate=-45]  {$x^{+} =2\pi \ $};
\draw (448.73,231.9) node [anchor=north west][inner sep=0.75pt]  [font=\LARGE]  {$2\pi $};
\draw (135,354.9) node [anchor=north west][inner sep=0.75pt]  [font=\LARGE]  {$-\pi $};
\draw (234.44,307) node  [font=\Large,rotate=-45]  {$x^{+} =0\ $};
\draw (386.94,302) node  [font=\Large,rotate=-315]  {$x^{-} =2\pi \ $};
\draw (233.03,147.88) node  [font=\Large,rotate=-315]  {$x^{-} =0\ $};

\end{tikzpicture}